\documentclass[prl,aps,showpacs,floatfix,floats,superscriptaddress,nobalancelastpage,twocolumn]{revtex4-1}
\usepackage{graphicx}
\usepackage{graphics}
\usepackage{latexsym,amsmath,amssymb,bm,euscript}
\usepackage{color}
\usepackage{dcolumn}
\usepackage{bm}
\usepackage{epstopdf}
\usepackage{times}
\usepackage{multirow}
\usepackage{wasysym}
\usepackage{subfigure}
\usepackage{bbm}

\def\ra{\rangle}
\def\la{\langle}

\begin{document}
\title{Kondo Destruction and Multipolar Order --  Implications for Heavy Fermion Quantum Criticality} 
\author{Hsin-Hua Lai}
\affiliation{Department of Physics and Astronomy, Rice University, Houston, Texas 77005, USA}
\author{Emilian M. Nica}
\affiliation{Department of Physics and Astronomy and Quantum Materials Institute,
University of British Columbia, Vancouver, B.C., V6T 1Z1, Canada}
\author{Wen-Jun Hu}
\affiliation{Department of Physics and Astronomy, Rice University, Houston, Texas 77005, USA}
\author{Shou-Shu Gong}
\affiliation{Department of Physics, Beihang University, Beijing 100191, China}
\author{Silke Paschen}
\affiliation{Institute of Solid State Physics, Vienna University of Technology,
Wiedner Hauptstr. 8-10, 1040 Vienna, Austria}
\author{Qimiao Si}
\affiliation{Department of Physics and Astronomy, Rice University, Houston, Texas 77005, USA}
\date{\today}

\begin{abstract}
Quantum criticality beyond the Landau paradigm represents a fundamental problem in condensed matter and statistical physics. 
Heavy fermion systems with multipolar degrees of freedom can play an important role in the search for its universal description. 
We consider a Kondo lattice model with both spin and quadrupole degrees of freedom,  which we show to exhibit 
an antiferroquadrupolar phase. Using  a field theoretical representation of the model, we find that Kondo couplings are 
exactly marginal in the renormalization group sense in this phase. This contrasts with the relevant nature of the Kondo couplings 
in the paramagnetic phase and, as such, it implies that a Kondo destruction and a concomitant small to large Fermi surface jump 
must occur as the system is tuned from the antiferroquadrupolar ordered to the paramagnetic phase.
Implications of our results for multipolar heavy fermion physics in particular and metallic quantum criticality in general are discussed.
\end{abstract}
\maketitle

\textit{Introduction}---
In strongly correlated systems, multiple building blocks often interplay with each other 
and create a variety of quantum phases and their transitions. 
Examples include the spin, orbital and nematic degrees of freedom in the iron-based systems~\cite{Kamihara2008,NatRevMat:2016},
which lead to a rich landscape of electronic orders, and the spin and valley degrees of freedom in bilayer graphenes twisted by magic angles~\cite{YCao1,YCao2}, which appear to yield a surprising Mott insulator near which superconductivity develops. 
The multiple degrees of freedom allow for not only the commonly observed antiferromagnetic (AF) states, 
but also ``hidden" orders, with unusual order parameters that  cannot readily be probed by experiments directly.
A prominent example is the quadrupolar order,  which breaks the spin-rotational symmetry 
as in any conventional magnetic order but, unlike the latter, preserves the time-reversal symmetry. Such an order has been proposed 
for frustrated magnetic systems~\cite{Papanicolaou1988,Tsunetsugu2006,Lauchli2006,Smerald2013} and even for 
the nematic phase of the iron-chalcogenide FeSe~\cite{YuSi_AFQ,QisiWang2016,Lai_AFQ2017}.
Multipolar degrees of freedom are also being discussed in noncollinear antiferromangets \cite{Suzuki:2017}.
They also arise in many heavy fermion metals, producing a variety of fascinating properties~\cite{Custers2012, Martelli2017, Sakai2011, Lee2018,Bauer2002,McCollam2013}.

Heavy fermion compounds typically involve local spin moments, which experience RKKY interactions between each other 
and Kondo interactions with conduction electrons, and exhibit quantum phase transitions between paramagnetic and AF ground states
\cite{SiSteglich2010,ColemanSchofield2005}. While the Kondo effect has been a hallmark of heavy fermion physics, 
a Kondo destruction has been shown to arise from the dynamical competition between the RKKY and Kondo interactions
~\cite{Si-Nature,Colemanetal}. It has been demonstrated in studies of Kondo lattice models from 
both the paramagnetic~\cite{Si-Nature} and AF-ordered ~\cite{Yamamoto2007} sides. Because the Kondo destruction yields
 quantum criticality that is beyond the Landau framework of order-parameter fluctuations,
it is important to assess its universality by considering settings that involve
other types of local degrees of freedom.

An especially opportune setting arises in heavy fermion systems with co-existing local spin and multipolar 
moments \cite{Custers2012,Martelli2017}, 
which allow for not only AF orders but also quadrupolar ones.
In Ce$_3$Pd$_{20}$Si$_6$, an antiferroquadrupolar (AFQ) order has been experimentally 
determined~\cite{Por16.1}, and a sequence of quantum critical points was discovered upon tuning by a magnetic field~\cite{Martelli2017}.
Theoretical calculations that approach the transitions from the paramagnetic side demonstrated a sequential Kondo destruction~\cite{Martelli2017}. This provides the motivation to study the Kondo effect and Fermi surface in the AFQ ordered state.
 
In this Letter, we address this pressing problem using a spin-$1$ Kondo lattice model, 
which contains both spin and quadrupole local degrees of freedom. We demonstrate a robust AFQ phase,
and describes its low-energy effective theory in terms of a quantum non-linear sigma model (NLsM) \cite{Smerald2013}.
Adapting a combined boson-fermion renormalization group (RG) procedure \cite{Yamamoto2010},
we show that the Kondo couplings are exactly marginal in the RG sense and thereby establish
a Kondo destruction in the multipolar order.

The 
model we consider is $H_{KL} = H_S + H_c + H_K$, with
\begin{eqnarray}
&& H_{S} = \sum_{i j}\left[ J_{ij}\left( \bm{S}_i \cdot \bm{S}_j \right) + K_{ij} \left(\bm{S}_i \cdot \bm{S}_j\right)^2\right], \label{Eq:BBQ}\\ 
&& H_c = \sum_{\bm{k}, \alpha = x,y,z} \epsilon_{\bm{k}} \psi_{\bm{k} \alpha}^\dagger \psi_{\bm{k} \alpha}, \\
 && H_K = \sum_j \left( J_K^I \bm{S}_j \cdot \bm{s}_{c,j} + J_K^{II} \bm{Q}_j \cdot \bm{q}_{c,j}\right), \label{Eq:Kondo}
\end{eqnarray}
where $H_S$ represents the spin-$1$ bilinear-biquadratic Hamiltonian,
in which we choose $J_{ij} = J_n,~K_{ij}= K_n $ for $i, j$ connected by $n$th-neighbor bonds. 
The spin-$1$ nature implies that the existence of local quadrupolar moments.
The 5-component quadrupolar operator 
at site $i$, ${\bf Q}_i$, can be defined as: $Q^{x^2 - y^2}_i = (S^x_i)^2 - (S^y_i)^2$, 
$Q^{3z^2 - r^2}_i = [ 2 (S^z_i)^2 - (S^x_i)^2 - (S^y_i)^2]/\sqrt{3}$, $Q^{xy}_i = S^x_i S^y_i + S^y_i S^x_i$, 
$Q^{yz}_i =  S^y_iS^z_i +S^z_i S^y_i$, and $Q^{zx}_i = S^z_i S^x_i + S^x_i S^z_i$. 
The biquadratic term can be re-expressed as $({\bf S}_i \cdot {\bf S}_j )^2 = ({\bf Q}_i \cdot {\bf Q}_j) /2 
- ({\bf S}_i \cdot {\bf S}_j) /2 + ({\bf S}^2_i {\bf S}^2_j)/3$. 
At the high-symmetry point, $J_n = K_n$, the symmetry is enhanced from SU(2) to SU(3).
Here, the spin and quadrupolar 
moments can be transformed to each other under SU(3) rotations. 
Our focus will be on the AFQ phase away from the SU(3) point; 
however, as we will see, the time-reversal-invariant 
basis 
that is natural for the SU(3) point -- 
which can be related to the  $s^z = \pm 1, 0$ basis under 
a unitary transformation -- will greatly facilitate our analysis. 
$H_c$ describes the conduction electrons,
which have three flavors with flavor index $\alpha = x, y, z$ in the SU(3) time-reversal-invariant basis. 
Within the 3-flavor conduction electron description, 
both the electrons' spin $\bm{s}_c$ and their $5$-component quadrupoles, $\bm{q}_c$, 
are expressed in bilinear forms (see Supplemental Materials for explicit forms of $\bm{q}_c$).
$H_K$ represents the Kondo couplings between the local moments and conduction electrons.  

\begin{figure}[t] 
   \includegraphics[width= 2.5 in]{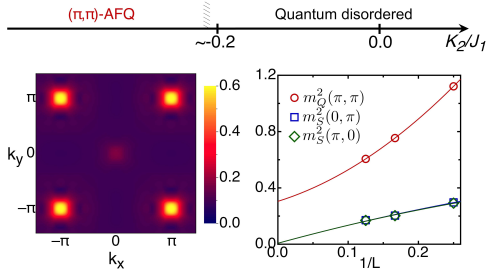} 
   \caption{(Color online) Top panel:Illustration of the phase diagram as a function of $K_2$ 
   that contains the $(\pi,\pi)$-AFQ with $J_1=1.0$ and $K_1=1.2$. 
   The quantum disordered phase has been studied before~\cite{Wenjun_NSL2017}.
Bottom left panel:The quadrupolar structure factor, $m_{\bm{Q}}$, which shows strong peaks at $(\pi, \pi)$. Bottom right panel:The finite-size scaling of the spin dipolar and spin quadrupolar order parameters, which shows finite $m_{\bm{Q}}$ at $\bm{q} = (\pi,\pi)$ and vanishing magnetic order parameter in the thermodynamic limit.}
   \label{Fig:DMRG_result}
\end{figure}

\textit{Existence of $(\pi,\pi)$ antiferroquadrupolar order}---
We first study the spin-$1$ bilinear-biquadratic lattice model, Eq.~\eqref{Eq:BBQ}, numerically using the large-scale Density Matrix Renormalization Group (DMRG) analysis.
It  is known that this 
SU(3)-symmetric point can host a phase with both spin and quadrupolar orders, 
which can be transformed to each other under SU(3) transformations
 \cite{Papanicolaou1988,Tsunetsugu2006,Lauchli2006,Smerald2013}. Away from the SU(3) point,
we find that increasing the weight of the biquadratic terms can 
stabilize the quadrupolar order. 
To illustrate 
 the robustness of the quadrupolar order in this model, we fix $J_1=1$, $K_1 = 1.2$, and 
  determine the phase diagram as a function of $K_2/J_1 (J_2 = 0)$,
which is shown in Fig.~\ref{Fig:DMRG_result}. We find the 
the AFQ order with $\bm{q}=(\pi,\pi)$.
This $(\pi,\pi)$-AFQ is a two-sublattice order characterized by the staggered expectation values of the $Q^{x^2 - y^2}$, \textit{i.e.}, 
$\langle Q^{x^2 - y^2}_j \rangle \sim (-1)^{j}$. 
The $(\pi,\pi)$-AFQ phase can be identified by calculating
the spin structure factor ($m^2_{\bm{S}}$) 
and quadrupolar structure factor ($m^2_{\bm{Q}}$).  As illustration, we consider the case $K_2/J_1 = - 0.3$,
which shows strong peak at $\bm{q} = (\pi,\pi)$ for $m^2_{\bm{Q}}$ and much weaker peak at $\bm{q} = (\pi,0)/(0,\pi)$
for $m^2_{\bm{S}}$, shown in the bottom left panel of Fig.~\ref{Fig:DMRG_result} for system size $L_y = 8$. 
Performing the finite-size scaling analysis, bottom right panel of Fig.~\ref{Fig:DMRG_result}, 
we find nonzero $m^2_{\bm{Q}}$ and vanishing $m^2_{\bm{S}}$, which 
shows the presence of the $(\pi,\pi)$-AFQ order.
\begin{figure}
\centering
 \subfigure[]{\label{Fig:pipiAFQ}\includegraphics[width= 1.2 in]{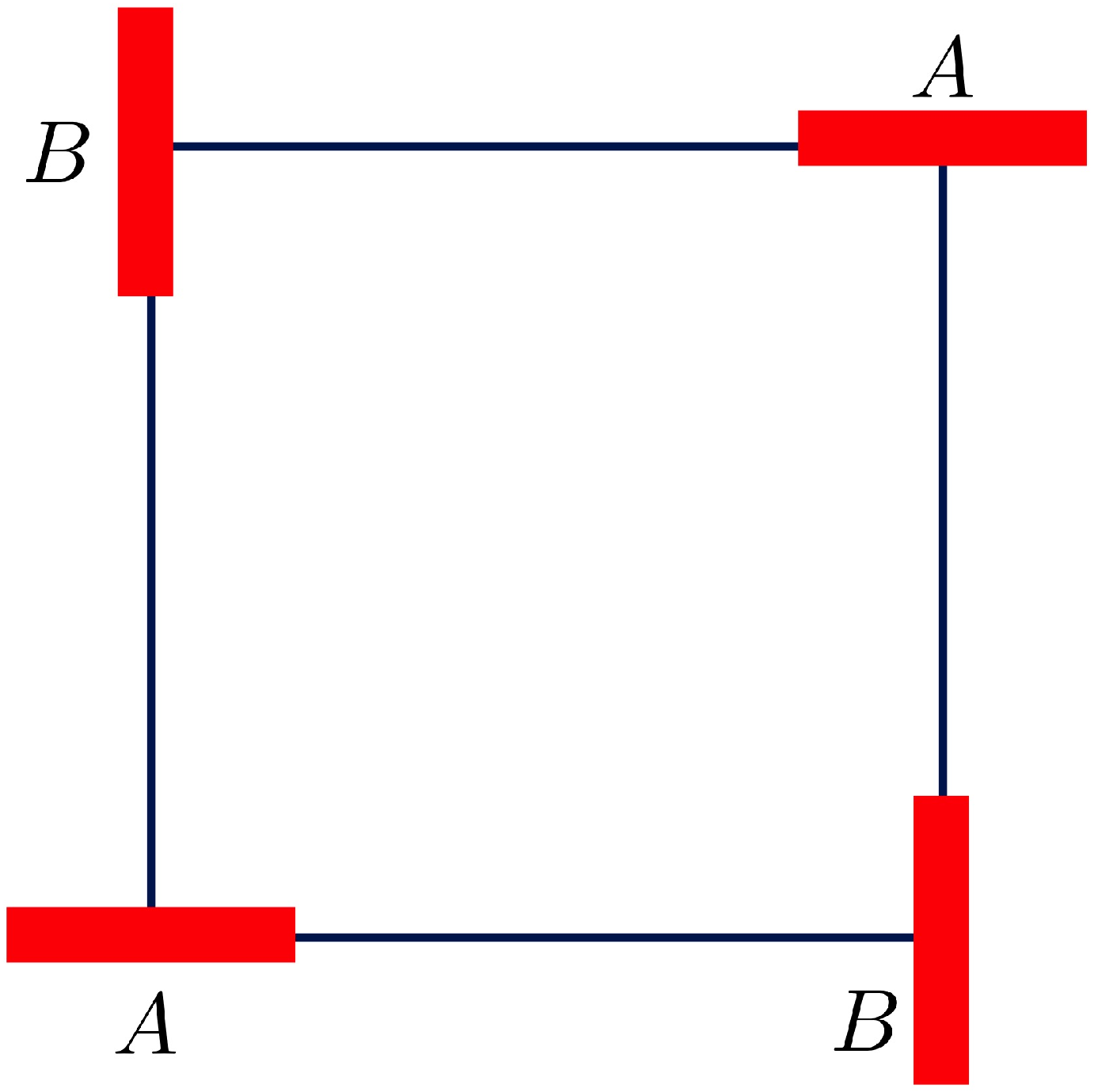}}~~~~
 \subfigure[]{\label{Fig:cluster_network}\includegraphics[width= 1.2 in]{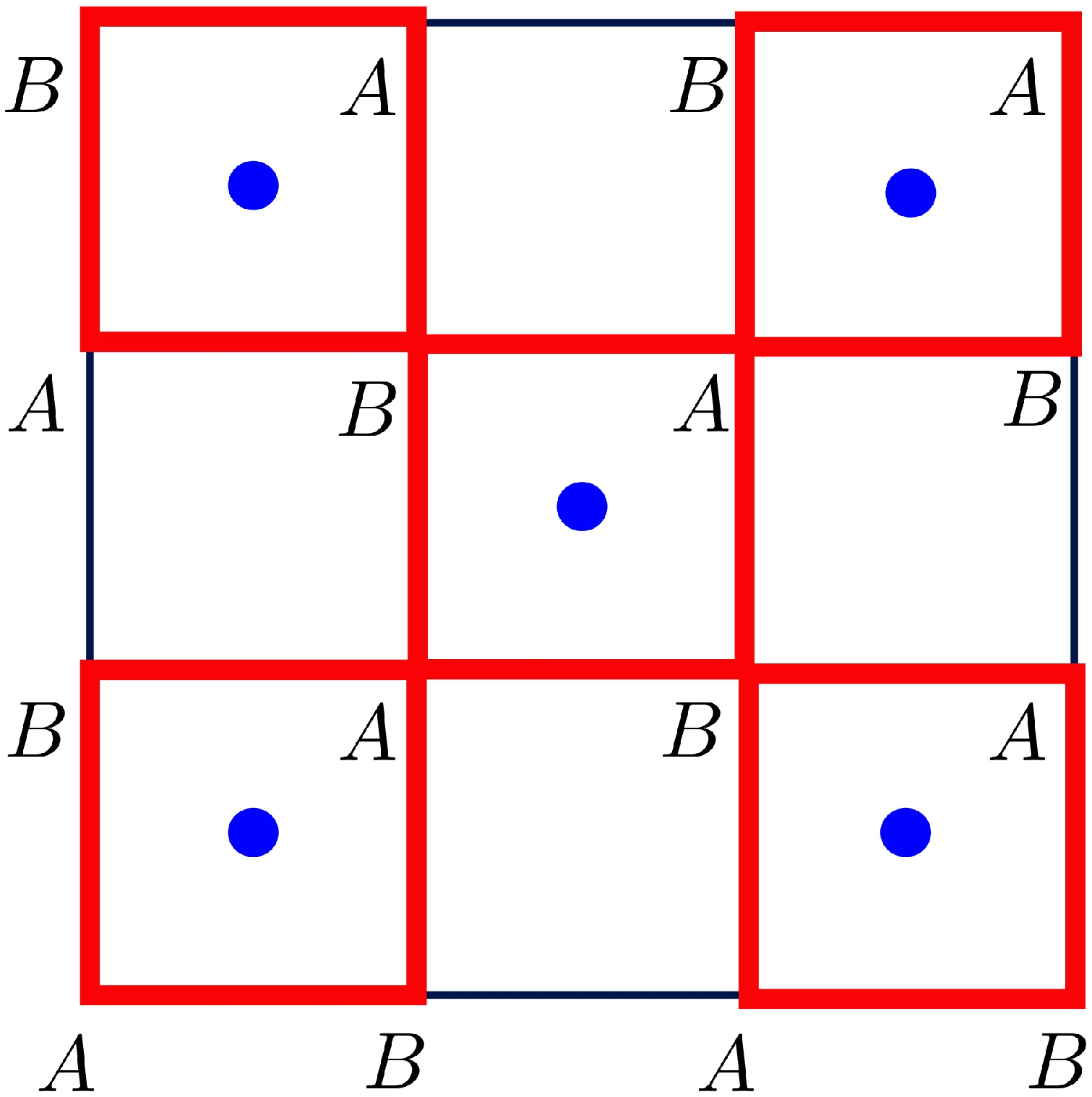}}~~~~
     \subfigure[]{\label{Fig:cluster_single}\includegraphics[width= 0.6 in]{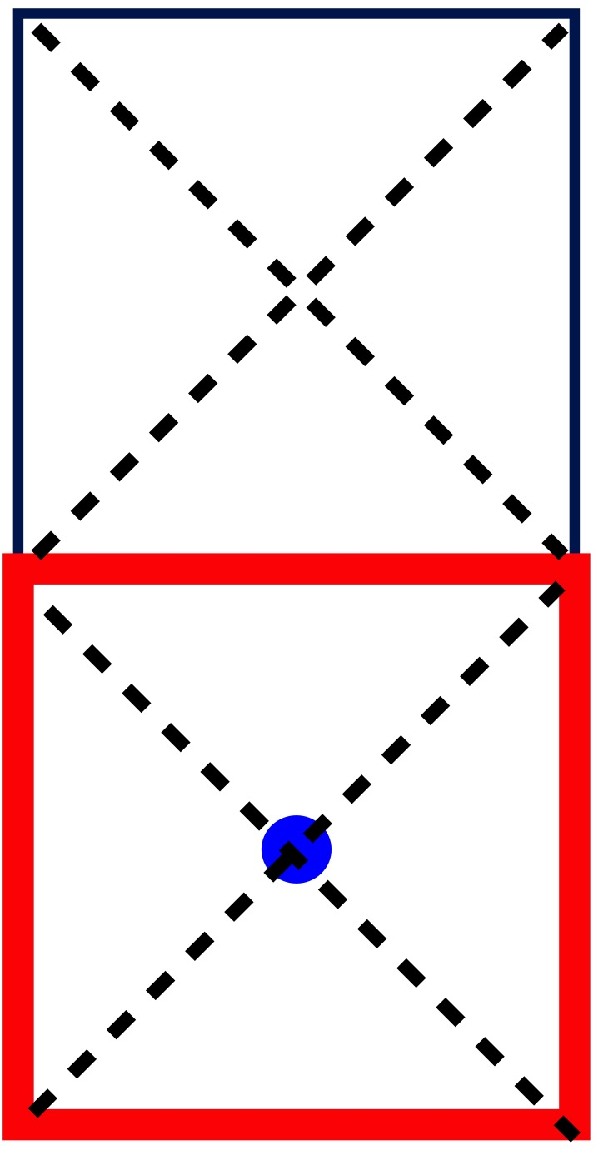}}
    \caption{(Color online) (a):The real-space pattern of $\bm{d}$ vectors for the lowest-energy state at SU(3) point 
    with $J_1=K_1 >0$ and $J_2 = K_2 <0$. (b):The partitioning of the square lattice used in the derivation of the field theory 
    for the $(\pi,\pi)$ AFQ order. The square lattice is divided into clusters (red squares) containing $8$ bonds. 
    Fields are defined at the centre of the clusters (blue dots), and we perform gradient expansions about these points. 
    (c):Illustration of a cluster containing 4 nearest-neighbor bonds (red lines) and 4 second-neighbor bonds (dashed lines).}
\label{Fig:cluster}
\end{figure}

\textit{Non-linear sigma model for $(\pi,\pi)$-AFQ}---
Because the commensurate AFQ breaks the spin-rotational symmetry but is time-reversal invariant, we can expect three 
Goldstone modes. To specify the low-energy effective theory including couplings that involve the conduction electrons, 
we 
describe  the Kondo lattice model 
using a NLsM representation by adapting the method illustrated in Ref.~\cite{Smerald2013}.
We first introduce the SU(3) time-reversal-invariant
 basis, $ |x\ra = i ( | 1 \ra - |\bar{1})/\sqrt{2},~~~ |y\ra = ( | 1\ra + | \bar{1} )/\sqrt{2}, ~~~|z \ra = -i | 0 \ra,$ where $|S^z = 1\ra \equiv |1\ra$.
 The
 state at site $j$ can be written 
 in terms of $ {\bf d}_j = (d^x_j,~d^y_j,~d^z_j)$, where $d_j^\alpha$ are complex numbers, with constraints
 from the normalization and from fixing the global phase among $d^{\alpha=x,y,a}$, \textit{i.e.}, 
 ${\bf d}_j \cdot \bar{\bf d}_j = 1$, and ${\bf d}_j^2 = \bar{\bf d}_j^2$, where $\bar{\bm{d}}$ means complex conjugate of ${\bm{d}}$.  
The Hamiltonian is then re-expressed as (see Supplemental Materials \cite{supp})
\begin{eqnarray}
H_{S} &=& \sum_{i j } \left[ J_n |{\bf d}_i \cdot \bar{\bf d}_j|^2 + J_n' | {\bf d}_i \cdot {\bf d}_j |^2  \right] \equiv H_{SU(3)} + H',~~~
\end{eqnarray}
where we have ignored the inconsequential constant terms.
We define the deviations from the SU(3) couplings, $J_n' \equiv K_n - J_n$,
and explicitly separate the SU(3)-invariant part of the Hamiltonian $H_{SU(3)}$ from the SU(3)-breaking part $H'$. 
At the SU(3) point, \textit{i.e.,} $J_n'=0$ or $J_n = K_n$, we can see that $H_S$ becomes a pure function 
of $|(\bm{d}_i \cdot \bar{\bm{d}}_j)|^2$. For the lowest-energy real-space pattern, we need to minimize 
the nearest-neighbor $|\bm{d}_i\cdot \bar{\bm{d}}_j|$ and maximize the $2$nd-neighbor $|\bm{d}_i \cdot \bar{\bm{d}}_j |$. 
For obtaining the NLsM description, we choose the ground state pattern of the $(\pi,\pi)$-AFQ, 
which satisfies the above requirement, as $ ({\bf d}^{gs}_A)^T = \begin{pmatrix} 1 & 0 & 0 \end{pmatrix}$, 
and $({\bf d}^{gs}_B)^T = \begin{pmatrix} 0 & 1 & 0 \end{pmatrix}$. 
Such a pattern is illustrated in Fig.~\ref{Fig:pipiAFQ}, which gives the correct $(\pi,\pi)$-AFQ order at the semi-classical level.

Starting from the SU(3) point, we know that the ground state energy is invariant under the global rotation ${\bf d} \rightarrow {\bf U}{\bf d}$, provided that ${\bf U}^{-1} = {\bf U}^\dagger$. To describe global rotations, we find that 
Gell-Mann matrices provide a natural choice of basis at the SU(3)-symmetric point. In general, we require $6$ distinct generators 
for SU(3). However, for the present $(\pi,\pi)$-AFQ phase, only $4$ out of $6$ are needed. 
The $4$ Gell-Mann matrices we choose are represented by $\lambda_{1- 4}$ (see Supplemental Materials for the explicit matrix forms). The global rotations in the complex space can be expressed as ${\bf U}(\phi) \equiv exp[{i\sum_{p =1}^4 \lambda_p \phi_p}].$ Besides the global rotations that preserve the ground state energy, we also need to consider the rotations involving the canting of the directors of ground state configurations, which increase the energy. The canting fields are represented as $\mu_{1 \sim 4}$ (see Supplemental Materials for details). 
The general rotations can be represented as
\begin{eqnarray}\label{Eq:rotation_m}
\nonumber {\bf D}(\phi, \ell, v) & = &  e^{i \sum_{p=1}^4 \lambda_p \phi_p + i \mu_1 \ell^z_1 + i \mu_2 v_A + i \mu_3 v_B + i \mu_4 \ell_2^z}\\
 && \hspace{-1.5 cm} \simeq U(\phi) \sum_{n=0}^\infty \left( i \mu_1 \ell^z_1 + i \mu_2 v_A + i \mu_3 v_B + i \mu_4 \ell_2^z\right)^n ,
\end{eqnarray}
where we approximately separate out the global rotation matrix $U(\phi)$ and Taylor expand the terms involving the canting fields. 
The general configuration of ${\bf d}_{a=A/B}$ can be obtained by applying the general rotation on the chosen ground state 
configurations ${\bf d}^{gs}_{a=A/B}$, \textit{i.e.}, $\bm{d}_a = \bm{D}(\phi,\ell,v) \bm{d}^{gs}_a$.
To obtain the low-energy descriptions within the harmonic theory, we keep the Taylor expansion up to $n=2$.
  Introducing $\ell^z \equiv \ell_1^z + i \ell^z_2$ and re-parametrizing
\begin{eqnarray}\label{Eq:U(n)}
{\bf U} = \begin{pmatrix}
n^x_A & n^x_B & n^x_C \\
n^y_A & n^y_B & n^y_C \\
n^z_A & n^z_B & n^z_C 
\end{pmatrix} = \begin{pmatrix}
\bm{n}_A & \bm{n}_B & \bm{n}_C
\end{pmatrix},
\end{eqnarray}
where the vectors ${\bf n}^T_a = \begin{pmatrix} n_a^x & n_a^y & n_a^z \end{pmatrix}$ inherit the constraints of ${\bf d}$ 
with ${\bf n}_a \cdot \bar{\bf n}_{b\not=a} = 0$, ${\bf n}_a^2 = {\bf n}_a^2$, and ${\bf n}_a \cdot \bar{\bf n}_a = 1$, with 
the vector ${\bf n}_c \equiv \bar{\bf n}_A \times \bar{\bf n}_B$ 
being introduced as a convenient piece of book-keeping in the present $(\pi, \pi)$-AFQ. It is 
not an  independent degrees of freedom, we can fully re-express $\bm{d}_{a = A/B}$ as functions of $\bm{n}_a$, 
$\ell^z$, and $v_{A/B}$ (see Supplemental Materials).  Taylor expanding $U(\phi)$ and keeping only the leading linear terms 
in $\phi$, we can see that $\bm{n}_a$ and the bosonic field $\phi_{1\sim 4}$ are related by
\begin{eqnarray} \label{Eq:n_to_phi}
\bm{n}_A \simeq \begin{pmatrix}
1 \\
-\phi_1 + i \phi_4 \\
\phi_2
\end{pmatrix};&~~
\bm{n}_B \simeq \begin{pmatrix}
\phi_1 + i \phi_4 \\
1 \\
-\phi_3
\end{pmatrix},
\end{eqnarray}
where 
 $\bm{n}_C = \bar{\bm{n}}_A \times \bar{\bm{n}}_B$. Away from the SU(3) point, we assume that $H'$ can be treated 
 perturbatively
 and does not affect our results in any significant manner.

Adopting the strategy of Ref.~\cite{Smerald2013}, we partition the square lattice into clusters
(Figs.~\ref{Fig:cluster_network}-\ref{Fig:cluster_single}),
 each of which containing 8 bonds [4 nearest-neighbor bonds and 4 second-neighbor bonds (dashed lines)],
 and perform the gradient expansion about the center of a cluster. Within the cluster picture, the partition function can be concisely expressed as $\mathcal{Z}_\square = \int D[{\bf d}] e^{- S_\square [{\bf d}]}$, where the action includes the kinetic terms $S_{kin}$ and the Hamiltonian terms $S_H$, $S_\square = S_{kin} + S_{Hs}$. The continuous descriptions of $S_{kin}=\int_0^\beta d\tau \mathcal{L}_{kin}$ and $S_{Hs}=\int_0^\beta d\tau H_S = \int_0^\beta d\tau \left[ H_{SU(3)} + H' \right]$ can be straightforwardly obtained after the gradient expansion. The detailed results are presented in the Supplemental Materials, and all the terms ($\mathcal{L}_{kin},~H_s$) are functions of 
 $\ell_1^z$, $\ell^z_2$, $v_{A/B}$, and $\bm{n}_{a} $. Integrating out the canting fields by solving the differential equations, $
\delta \mathcal{L}/\delta \ell_1^z = 0$, $\delta \mathcal{L}/ \delta \ell_2^z =0$, $\delta \mathcal{L}/\delta v_A = 0$, 
$\delta \mathcal{L}/ \delta v_B =0$ within the steepest-descent approximation, we obtain the NLsM for the $(\pi,\pi)$-AFQ 
at the harmonic level. 
Stability requires 
 $K_1>0$, $ K_2 <0$, $K_1 - J_1 \geq 0$, $J_2 - K_2 >0$, 
$J_1 + J_2 - K_2 > 0$, and $K_1 + 2K_2 - 4J_2 \geq 0$. 
Focusing on the regime away from the SU(3) point, $J_n' \not=0$, we find that $\phi_4$ can be ignored 
since it represents the spin-wave mode that is always gapped due to the finite mass term. 
The effective NLsM for the $(\pi,\pi)$-AFQ is
\begin{eqnarray}\label{Eq:Ls}
\nonumber \mathcal{L}^{NLsM}_S \simeq && \frac{ ( \partial_\tau \phi_1 )^2}{4(J_1 + J_2 - K_2)} + 2 (K_1 - 2 K_2) \sum_\lambda (\partial_\lambda \phi_1)^2  \\
&& + \sum_{a=2,3} \left[ \frac{(\partial_\tau \phi_a)^2}{8( J_2 - K_2)} - 2K_2 \sum_\lambda (\partial_\lambda \phi_a)^2\right]. 
\end{eqnarray}
%

\textit{Kondo couplings}--- 
Using the identity $\bm{d}_a = D(\phi,\ell,v) \bm{d}^{gs}_a$, and Eqs.~\eqref{Eq:rotation_m}-\eqref{Eq:n_to_phi}, 
we can straightforwardly write down the fluctuating $8$-component spin/quadrupolar field in the NLsM description (detailed in Supplemental Materials). Concisely, the $8$-component field 
can be separate into a uniform part and an oscillating part,
\begin{eqnarray}
Q(\bm{r}) \simeq \mathcal{Q}_{\bm{0}} + \mathcal{Q}_{\bm{M}} \cos(\bm{M}\cdot \bm{r}),~~~~~
\end{eqnarray}
where $\bm{M} \equiv (\pi,\pi)$, $\mathcal{Q}_{\bm{q}}$ represents the low-energy $8$-component field with momenta $\bm{q}=\bm{0}, \bm{M}$. We remark that that the uniform part contains a ``static" background of $Q^{3z^2 -r^2}$ that can directly couple to the $q_c^{3z^2 -r^2}$ of the $3$-flavor conduction electrons due to he Kondo couplings, Eq.~\eqref{Eq:Kondo}. This static background $Q^{3z^2 - r^2}$ field is only invariant under rotation between $x$-$y$ plane, which breaks the SU(3) symmetry of the conduction elections down to SU(2)$\times$U(1), where SU(2) is spanned by the $c_x$ and $c_y$ conduction elections and U(1) is spanned by $c_z$. We expect that 
the Fermi velocities of  $c_x$ and $c_y$ are the same ($v_x = v_y$) but is different from that of $c_z~(v_z)$. 

To be specific, at low energies, the conduction electrons in the presence of the static quadrupolar background
is 
\begin{eqnarray}\label{Eq:Sc}
\nonumber S_c && \simeq  \int d^d\bm{K} d\epsilon \sum_{\alpha = x,y} \psi^\dagger_\alpha (\bm{K}, i \epsilon) ( i \epsilon - \xi_K)\psi_\alpha(\bm{K},i\epsilon) + \\
&& + \int d^d\bm{K}' d\epsilon' \psi^\dagger_z(\bm{K}',i\epsilon')(i\epsilon' - \tilde{\xi}_{K'}) \psi_z(\bm{K}',i\epsilon'),
\end{eqnarray}
where $\xi_K = v_F(K - K_F)$, and $\tilde{\xi}_{K'} = \tilde{v}_{F}(K' - K'_F)$, where $K_F$ and $K'_F$ are Fermi momenta for $c_{x/y}$ and $c_z$ and are generically different. 
The  spin dipolar and quadrupolar degrees of freedom consisting of $\psi_{x/y}$ and $\psi_{z}$ fermions, $e.g.$, 
$s^x_c \sim \psi^\dagger_y \psi_z - \psi^\dagger_z \psi_y$, can be ignored due to the finite energy gap $(\Delta E)$ 
between the $\psi_{x/y}$ bands and the $\psi_{z}$ band, $\Delta E \propto J_K \la Q^{3z^2 - r^2} \ra$.

Shifting our focus to the Kondo couplings, Eq.~\eqref{Eq:Kondo}, we can now re-express them as 
\begin{eqnarray}
\nonumber \mathcal{L}_K = && -J_K^I   \left( 2\ell^z_2 + \frac{i}{2} \left(n^y_A - \bar{n}^y_A - n^x_B  + \bar{n}^x_B\right)\right)s_c^z - \\
&&  - J_K^{II}  \left( 2 \ell^z_1 + \frac{1}{2}\left( n^y_A + \bar{n}^y_A + n^x_B + \bar{n}^x_B \right)\right) q_c^{xy},~~~
\end{eqnarray}
where only the uniform part $\mathcal{Q}_{\bm{0}}$ couples to the conduction electrons
near the Fermi surface.
We integrate out the canting field within the steepest descent approximation.
After some algebra, we conclude that the effective low-energy description of the action 
is $S = S^{NLsM}_s + S_K + S_c$, where $S^{NLsM}_S$ and $S_c$ are defined in Eqs.~\eqref{Eq:Ls} and \eqref{Eq:Sc}, and $S_K$ is
\begin{eqnarray}
S_K & \simeq & \int_0^\beta \frac{d\tau}{2} \int d^2 r  \lambda_z \left( s^z_c \partial_\tau \phi_1 \right),
\end{eqnarray}
where $\lambda_z = -i J_K^I/[4(J_1 + J_2 - K_2 )]$ is the dimensionless coupling. 

\textit{Exact marginality of Kondo couplings}---
We now analyze the scaling of the Kondo coupling $\lambda_z$ in the $(\pi,\pi)$-AFQ using the RG procedure 
described in Ref.~\cite{Yamamoto2010}.
For clarity, we introduce $\Phi_a(\bm{r},\tau) \equiv \partial_\tau \phi_a(\bm{r},\tau)$. 
The scaling dimension of $\Phi_a(\bm{r},\tau)$ can be directly read out, $\Delta[\Phi_a(\bm{r},\tau)] = 1$, 
indicating that
the scaling dimension of its Fourier partner as $\Delta[\Phi_a (\bm{q},\omega)] = -d$, where $d$ is the spatial dimension. 
For the conduction electron fields, we obtain that $\Delta[\psi_c(\bm{K}, \omega)] = -3/2$. 
We can see that, at the tree level, 
the Kondo coupling is \textit{marginal}, $\Delta[S_K] = \Delta[ dk d\epsilon d^dq d \omega \psi^{x \dagger}_c (\bm{k} + \bm{q}, \epsilon + \omega) \psi^y_c (\bm{k}, \omega) \Phi_1 (\bm{q},\omega)] = 1 + 1 + d + 1 + 2 (-3/2) -d = 0$ (see Supplemental Materials). .
 
We then  turn to what happens beyond the tree level.
Considering a spherical Fermi surface of conduction electrons, we
approximate their contribution via a momentum integral near Fermi surface.
Keeping the most relevant term, we obtain $\int d^d \bm{K} = \int^{K_F + \Lambda}_{K_F - \Lambda} K^{d-1} dK \int d^{d-1} \Omega_K \simeq  K_F^{d-1} \int_{-\Lambda}^\Lambda dk \int d^{d-1} \Omega_K$, where we introduce $ k = K_F - K$ and keep only the $K_F^{d-1}$ terms after Taylor expansion. Now the kinetic part of the fermions can be re-expressed as
\begin{eqnarray}
\nonumber  S_c &\simeq& K_F^{d-1} \int dk_\alpha d^{d-1} \Omega_K d\epsilon \psi^{\alpha \dagger}_c \left( i \epsilon_\alpha - v_F^\alpha k_\alpha \right) \psi^\alpha_c \\
 &=& \int d\bar{k}_\alpha d^{d-1}\bar{\Omega}_K d\bar{\epsilon} \bar{\psi}_c^{\alpha \dagger} \left ( \bar{\epsilon}_\alpha - v_F^\alpha \bar{k}_\alpha \right) \bar{\psi}_c^\alpha,
\end{eqnarray}
where we introduce the dimensionless couplings, $\epsilon = \Lambda \bar{\epsilon}$, $k = \Lambda \bar{k}$, $\Omega_K = \bar{\Omega}_K$, $K_F^{d-1} \Lambda^3 \psi^\dagger \psi = \bar{\psi}^\dagger \bar{\psi}$. For the action of the bosonic fields, Eq.~\eqref{Eq:Ls}, we perform similar transformation, $\int d^2\bm{q} d\omega (\bm{q}^2 + \omega^2) \phi_1^2(\bm{q},\omega) = \int d^2\bar{\bm{q}} d \bar{\omega} ( \bar{\bm{q}}^2 + \bar{\omega}^2 ) \bar{\phi}_1^2 (\bar{\bm{q}},\bar{\omega})$, where we define $\Lambda^5 \phi_1^2 = \bar{\phi}_1^2$. Plugging the new definition into the Kondo action, we find that at $d=2$ it takes the form $\sim \int dk d\epsilon d^2q d\omega \psi_c^{x \dagger} (k+q, \epsilon +\omega)   \psi_c^y (k,\omega)  \omega  \phi_1 (q, \omega)$, which leads to
\begin{eqnarray}
\frac{S_K}{S_c} \propto \frac{\sqrt{\Lambda}}{K_F},
\end{eqnarray}
where we can see in the limit $\Lambda/K_F \rightarrow 0$, \textit{i.e.}, the Fermi momentum is much larger than the thin-shell momentum cut-off near the Fermi surface, the Kondo coupling is heavily suppressed. Therefore, the Kondo vertex is associated with positive powers of $\sqrt{\Lambda}/K_F$ which is vanishingly small. As the number of powers of Kondo couplings increases, so does the suppression factor, and, therefore, all higher-order terms are suppressed, which means that the scaling result at tree-level RG analysis is exact. The Kondo coupling is indeed exactly \textit{marginal}.

This exact marginality implies that the Kondo coupling does not flow to strong coupling. In other words, 
in the AFQ phase, the local 
moments do not form a multipolar Kondo singlet with the conduction electrons. 
Thus, the RG fixed point in the parameter regime we consider, namely weak Kondo coupling in the presence of an AFQ, 
shows the physics of Kondo destruction.

\textit{Implications for the quantum phases and their transitions in heavy fermion metals}---
The Kondo destruction we have shown, when the multipolar RKKY interactions dominate over the corresponding Kondo interactions, 
has a clear physical picture. In the AFQ order, the local degrees of freedom are strongly coupled with each other and 
become manifested as three quadrupolar Goldstone modes at low energies. 
Because these are collective bosonic modes, they can 
scatter the conduction electrons, but cannot form an entangled Kondo singlet with the latter. 
By contrast, it is well-known that when the Kondo interactions dominate over the RKKY interactions, they are marginally relevant and 
flow towards strong coupling, thereby yielding a Kondo entangled state~\onlinecite{Martelli2017}; physically, the local degrees of freedom 
will be able to lower the ground state energy of the system by binding with the conduction electrons into a singlet state. 
Calculations on the dynamical competition between RKKY and Kondo interactions
from the Kondo-dominated side in Ref.~\onlinecite{Martelli2017} led to the proposal for two stages of Kondo destructions.
Our asymptotically exact results from the opposite end shows that 
multipolar Kondo destruction does take place on the ordered side. As such, our results help establish a robust theoretical foundation
for the notion of sequential Kondo destruction~\cite{Martelli2017}.

Our findings set the stage for detailed studies of 
 heavy fermion materials with both spin and orbital moments in their ground state.
The simplest case arises in Ce-based systems of cubic point symmetry if the $\Gamma_8$ quartet is the ground state of the 
$^2\rm{F}_{5/2}$ multiplet \cite{Pas14.1}. Examples where a continuous phase transition to a state with AFQ order was observed are,
in addition to the aforementioned Ce$_3$Pd$_{20}$Si$_6$ \cite{Por16.1}, CeB$_6$ \cite{Nak01.1} and, 
tentatively, CeTe~\cite{Kaw11.1} and CeCoSi \cite{Tan18.1} under pressure.

\textit{Conclusion}--
We have studied a spin-$1$ Kondo lattice model with co-existing spin and quadrupolar local moments and used
density matrix renormalization group analysis to firmly 
demonstrate the presence of an antiferroquadrupolar order.
We have  derived a non-linear sigma model description of the antiferroquadrupolar order and,
based on a renormalization-group analysis, found  that the Kondo couplings are exactly marginal in this phase.
Our results help provide a robust theoretical foundation for the recently advanced 
notion of sequential localization in multipolar Kondo lattice systems
~\cite{Martelli2017}.
Our findings point to a growing list of heavy fermion metals with multipolar degrees of freedom as a new setting
for the exploration towards a universal description of beyond-Landau quantum criticality and strange metal physics.
In general, they illustrate how the interplay between entwined degrees of freedom can give rise to novel phases and unusual excitations, 
a theme that is centrally important to a broad range of strongly correlated systems.

\textit{Acknowledgement}--
The work at Rice was in part supported
by the NSF (DMR-1611392), the Robert A. Welch Foundation (C-1411),
the ARO (W911NF-14-1-0525), the Big-Data
Private-Cloud Research Cyberinfrastructure MRI Award funded by NSF (CNS-1338099),
and an IBM Shared University Research (SUR) Award.
H.-H.L. has been supported by a Smalley Postdoctoral
Fellowship at the Rice Center for Quantum Materials.
SP acknowledges financial support from the Austrian Science Fund (project
P29296-N27).
Q.S. acknowledges the support of
ICAM and a QuantEmX grant from the Gordon and Betty Moore
Foundation through Grant No. GBMF5305, 
the hospitality
of University of California
at Berkeley and of the Aspen Center for Physics, which is supported by NSF grant
No. PHY-1607611,
and the hospitality  and support by a Ulam Scholarship 
from the Center for Nonlinear Studies at Los Alamos National Laboratory.
\bibliography{biblio4AFQ}

\begin{thebibliography}{31}%
\makeatletter
\providecommand \@ifxundefined [1]{%
 \@ifx{#1\undefined}
}%
\providecommand \@ifnum [1]{%
 \ifnum #1\expandafter \@firstoftwo
 \else \expandafter \@secondoftwo
 \fi
}%
\providecommand \@ifx [1]{%
 \ifx #1\expandafter \@firstoftwo
 \else \expandafter \@secondoftwo
 \fi
}%
\providecommand \natexlab [1]{#1}%
\providecommand \enquote  [1]{``#1''}%
\providecommand \bibnamefont  [1]{#1}%
\providecommand \bibfnamefont [1]{#1}%
\providecommand \citenamefont [1]{#1}%
\providecommand \href@noop [0]{\@secondoftwo}%
\providecommand \href [0]{\begingroup \@sanitize@url \@href}%
\providecommand \@href[1]{\@@startlink{#1}\@@href}%
\providecommand \@@href[1]{\endgroup#1\@@endlink}%
\providecommand \@sanitize@url [0]{\catcode `\\12\catcode `\$12\catcode
  `\&12\catcode `\#12\catcode `\^12\catcode `\_12\catcode `\%12\relax}%
\providecommand \@@startlink[1]{}%
\providecommand \@@endlink[0]{}%
\providecommand \url  [0]{\begingroup\@sanitize@url \@url }%
\providecommand \@url [1]{\endgroup\@href {#1}{\urlprefix }}%
\providecommand \urlprefix  [0]{URL }%
\providecommand \Eprint [0]{\href }%
\providecommand \doibase [0]{http://dx.doi.org/}%
\providecommand \selectlanguage [0]{\@gobble}%
\providecommand \bibinfo  [0]{\@secondoftwo}%
\providecommand \bibfield  [0]{\@secondoftwo}%
\providecommand \translation [1]{[#1]}%
\providecommand \BibitemOpen [0]{}%
\providecommand \bibitemStop [0]{}%
\providecommand \bibitemNoStop [0]{.\EOS\space}%
\providecommand \EOS [0]{\spacefactor3000\relax}%
\providecommand \BibitemShut  [1]{\csname bibitem#1\endcsname}%
\let\auto@bib@innerbib\@empty
\bibitem [{\citenamefont {Kamihara}\ \emph {et~al.}(2008)\citenamefont
  {Kamihara}, \citenamefont {Watanabe}, \citenamefont {Hirano},\ and\
  \citenamefont {Hosono}}]{Kamihara2008}%
  \BibitemOpen
  \bibfield  {author} {\bibinfo {author} {\bibfnamefont {Y.}~\bibnamefont
  {Kamihara}}, \bibinfo {author} {\bibfnamefont {T.}~\bibnamefont {Watanabe}},
  \bibinfo {author} {\bibfnamefont {M.}~\bibnamefont {Hirano}}, \ and\ \bibinfo
  {author} {\bibfnamefont {H.}~\bibnamefont {Hosono}},\ }\href {\doibase
  10.1021/ja800073m} {\bibfield  {journal} {\bibinfo  {journal} {{Journal of
  the American Chemical Society}}\ }\textbf {\bibinfo {volume} {130}},\
  \bibinfo {pages} {3296} (\bibinfo {year} {{2008}})}\BibitemShut {NoStop}%
\bibitem [{\citenamefont {Si}\ \emph {et~al.}(2016)\citenamefont {Si},
  \citenamefont {Yu},\ and\ \citenamefont {Abrahams}}]{NatRevMat:2016}%
  \BibitemOpen
  \bibfield  {author} {\bibinfo {author} {\bibfnamefont {Q.}~\bibnamefont
  {Si}}, \bibinfo {author} {\bibfnamefont {R.}~\bibnamefont {Yu}}, \ and\
  \bibinfo {author} {\bibfnamefont {E.}~\bibnamefont {Abrahams}},\ }\href
  {\doibase 10.1038/natrevmats.2016.17} {\bibfield  {journal} {\bibinfo
  {journal} {Nature Reviews Materials}\ }\textbf {\bibinfo {volume} {1}},\
  \bibinfo {pages} {16017} (\bibinfo {year} {2016})}\BibitemShut {NoStop}%
\bibitem [{\citenamefont {Cao}\ \emph {et~al.}(2018{\natexlab{a}})\citenamefont
  {Cao}, \citenamefont {Fatemi}, \citenamefont {Fang}, \citenamefont
  {Watanabe}, \citenamefont {Taniguchi}, \citenamefont {Kaxiras},\ and\
  \citenamefont {Jarillo-Herrero}}]{YCao1}%
  \BibitemOpen
  \bibfield  {author} {\bibinfo {author} {\bibfnamefont {Y.}~\bibnamefont
  {Cao}}, \bibinfo {author} {\bibfnamefont {V.}~\bibnamefont {Fatemi}},
  \bibinfo {author} {\bibfnamefont {S.}~\bibnamefont {Fang}}, \bibinfo {author}
  {\bibfnamefont {K.}~\bibnamefont {Watanabe}}, \bibinfo {author}
  {\bibfnamefont {T.}~\bibnamefont {Taniguchi}}, \bibinfo {author}
  {\bibfnamefont {E.}~\bibnamefont {Kaxiras}}, \ and\ \bibinfo {author}
  {\bibfnamefont {P.}~\bibnamefont {Jarillo-Herrero}},\ }\href {\doibase
  10.1038/nature26160} {\bibfield  {journal} {\bibinfo  {journal} {Nature}\
  }\textbf {\bibinfo {volume} {556}},\ \bibinfo {pages} {43} (\bibinfo {year}
  {2018}{\natexlab{a}})}\BibitemShut {NoStop}%
\bibitem [{\citenamefont {Cao}\ \emph {et~al.}(2018{\natexlab{b}})\citenamefont
  {Cao}, \citenamefont {Fatemi}, \citenamefont {Demir}, \citenamefont {Fang},
  \citenamefont {Tomarken}, \citenamefont {Luo}, \citenamefont
  {Sanchez-Yamagishi}, \citenamefont {Watanabe}, \citenamefont {Taniguchi},
  \citenamefont {Kaxiras}, \citenamefont {Ashoori},\ and\ \citenamefont
  {Jarillo-Herrero}}]{YCao2}%
  \BibitemOpen
  \bibfield  {author} {\bibinfo {author} {\bibfnamefont {Y.}~\bibnamefont
  {Cao}}, \bibinfo {author} {\bibfnamefont {V.}~\bibnamefont {Fatemi}},
  \bibinfo {author} {\bibfnamefont {A.}~\bibnamefont {Demir}}, \bibinfo
  {author} {\bibfnamefont {S.}~\bibnamefont {Fang}}, \bibinfo {author}
  {\bibfnamefont {S.~L.}\ \bibnamefont {Tomarken}}, \bibinfo {author}
  {\bibfnamefont {J.~Y.}\ \bibnamefont {Luo}}, \bibinfo {author} {\bibfnamefont
  {J.~D.}\ \bibnamefont {Sanchez-Yamagishi}}, \bibinfo {author} {\bibfnamefont
  {K.}~\bibnamefont {Watanabe}}, \bibinfo {author} {\bibfnamefont
  {T.}~\bibnamefont {Taniguchi}}, \bibinfo {author} {\bibfnamefont
  {E.}~\bibnamefont {Kaxiras}}, \bibinfo {author} {\bibfnamefont {R.~C.}\
  \bibnamefont {Ashoori}}, \ and\ \bibinfo {author} {\bibfnamefont
  {P.}~\bibnamefont {Jarillo-Herrero}},\ }\href {\doibase 10.1038/nature26154}
  {\bibfield  {journal} {\bibinfo  {journal} {Nature}\ }\textbf {\bibinfo
  {volume} {556}},\ \bibinfo {pages} {80} (\bibinfo {year}
  {2018}{\natexlab{b}})}\BibitemShut {NoStop}%
\bibitem [{\citenamefont {Papanicolaou}(1988)}]{Papanicolaou1988}%
  \BibitemOpen
  \bibfield  {author} {\bibinfo {author} {\bibfnamefont {N.}~\bibnamefont
  {Papanicolaou}},\ }\href {\doibase 10.1016/0550-3213(88)90073-9} {\bibfield
  {journal} {\bibinfo  {journal} {Nuclear Physics B}\ }\textbf {\bibinfo
  {volume} {305}},\ \bibinfo {pages} {367 } (\bibinfo {year}
  {1988})}\BibitemShut {NoStop}%
\bibitem [{\citenamefont {Tsunetsugu}\ and\ \citenamefont
  {Arikawa}(2006)}]{Tsunetsugu2006}%
  \BibitemOpen
  \bibfield  {author} {\bibinfo {author} {\bibfnamefont {H.}~\bibnamefont
  {Tsunetsugu}}\ and\ \bibinfo {author} {\bibfnamefont {M.}~\bibnamefont
  {Arikawa}},\ }\href {\doibase 10.1143/JPSJ.75.083701} {\bibfield  {journal}
  {\bibinfo  {journal} {Journal of the Physical Society of Japan}\ }\textbf
  {\bibinfo {volume} {75}},\ \bibinfo {pages} {083701} (\bibinfo {year}
  {2006})}\BibitemShut {NoStop}%
\bibitem [{\citenamefont {L\"auchli}\ \emph {et~al.}(2006)\citenamefont
  {L\"auchli}, \citenamefont {Mila},\ and\ \citenamefont {Penc}}]{Lauchli2006}%
  \BibitemOpen
  \bibfield  {author} {\bibinfo {author} {\bibfnamefont {A.}~\bibnamefont
  {L\"auchli}}, \bibinfo {author} {\bibfnamefont {F.}~\bibnamefont {Mila}}, \
  and\ \bibinfo {author} {\bibfnamefont {K.}~\bibnamefont {Penc}},\ }\href
  {\doibase 10.1103/PhysRevLett.97.087205} {\bibfield  {journal} {\bibinfo
  {journal} {Phys. Rev. Lett.}\ }\textbf {\bibinfo {volume} {97}},\ \bibinfo
  {pages} {087205} (\bibinfo {year} {2006})}\BibitemShut {NoStop}%
\bibitem [{\citenamefont {Smerald}\ and\ \citenamefont
  {Shannon}(2013)}]{Smerald2013}%
  \BibitemOpen
  \bibfield  {author} {\bibinfo {author} {\bibfnamefont {A.}~\bibnamefont
  {Smerald}}\ and\ \bibinfo {author} {\bibfnamefont {N.}~\bibnamefont
  {Shannon}},\ }\href {\doibase 10.1103/PhysRevB.88.184430} {\bibfield
  {journal} {\bibinfo  {journal} {Phys. Rev. B}\ }\textbf {\bibinfo {volume}
  {88}},\ \bibinfo {pages} {184430} (\bibinfo {year} {2013})}\BibitemShut
  {NoStop}%
\bibitem [{\citenamefont {Yu}\ and\ \citenamefont {Si}(2015)}]{YuSi_AFQ}%
  \BibitemOpen
  \bibfield  {author} {\bibinfo {author} {\bibfnamefont {R.}~\bibnamefont
  {Yu}}\ and\ \bibinfo {author} {\bibfnamefont {Q.}~\bibnamefont {Si}},\ }\href
  {\doibase 10.1103/PhysRevLett.115.116401} {\bibfield  {journal} {\bibinfo
  {journal} {Phys. Rev. Lett.}\ }\textbf {\bibinfo {volume} {115}},\ \bibinfo
  {pages} {116401} (\bibinfo {year} {2015})}\BibitemShut {NoStop}%
\bibitem [{\citenamefont {Wang}\ \emph {et~al.}(2016)\citenamefont {Wang},
  \citenamefont {Shen}, \citenamefont {Pan}, \citenamefont {Zhang},
  \citenamefont {Ikeuchi}, \citenamefont {Iida}, \citenamefont {Christianson},
  \citenamefont {Walker}, \citenamefont {Adroja}, \citenamefont {Abdel-Hafiez},
  \citenamefont {Chen}, \citenamefont {Chareev}, \citenamefont {Vasiliev},\
  and\ \citenamefont {Zhao}}]{QisiWang2016}%
  \BibitemOpen
  \bibfield  {author} {\bibinfo {author} {\bibfnamefont {Q.}~\bibnamefont
  {Wang}}, \bibinfo {author} {\bibfnamefont {Y.}~\bibnamefont {Shen}}, \bibinfo
  {author} {\bibfnamefont {B.}~\bibnamefont {Pan}}, \bibinfo {author}
  {\bibfnamefont {X.}~\bibnamefont {Zhang}}, \bibinfo {author} {\bibfnamefont
  {K.}~\bibnamefont {Ikeuchi}}, \bibinfo {author} {\bibfnamefont
  {K.}~\bibnamefont {Iida}}, \bibinfo {author} {\bibfnamefont {A.~D.}\
  \bibnamefont {Christianson}}, \bibinfo {author} {\bibfnamefont {H.~C.}\
  \bibnamefont {Walker}}, \bibinfo {author} {\bibfnamefont {D.~T.}\
  \bibnamefont {Adroja}}, \bibinfo {author} {\bibfnamefont {M.}~\bibnamefont
  {Abdel-Hafiez}}, \bibinfo {author} {\bibfnamefont {X.}~\bibnamefont {Chen}},
  \bibinfo {author} {\bibfnamefont {D.~A.}\ \bibnamefont {Chareev}}, \bibinfo
  {author} {\bibfnamefont {A.~N.}\ \bibnamefont {Vasiliev}}, \ and\ \bibinfo
  {author} {\bibfnamefont {J.}~\bibnamefont {Zhao}},\ }\href@noop {} {\bibfield
   {journal} {\bibinfo  {journal} {Nature Communications}\ }\textbf {\bibinfo
  {volume} {7}},\ \bibinfo {pages} {12182} (\bibinfo {year}
  {2016})}\BibitemShut {NoStop}%
\bibitem [{\citenamefont {Lai}\ \emph {et~al.}(2017)\citenamefont {Lai},
  \citenamefont {Hu}, \citenamefont {Nica}, \citenamefont {Yu},\ and\
  \citenamefont {Si}}]{Lai_AFQ2017}%
  \BibitemOpen
  \bibfield  {author} {\bibinfo {author} {\bibfnamefont {H.-H.}\ \bibnamefont
  {Lai}}, \bibinfo {author} {\bibfnamefont {W.-J.}\ \bibnamefont {Hu}},
  \bibinfo {author} {\bibfnamefont {E.~M.}\ \bibnamefont {Nica}}, \bibinfo
  {author} {\bibfnamefont {R.}~\bibnamefont {Yu}}, \ and\ \bibinfo {author}
  {\bibfnamefont {Q.}~\bibnamefont {Si}},\ }\href {\doibase
  10.1103/PhysRevLett.118.176401} {\bibfield  {journal} {\bibinfo  {journal}
  {Phys. Rev. Lett.}\ }\textbf {\bibinfo {volume} {118}},\ \bibinfo {pages}
  {176401} (\bibinfo {year} {2017})}\BibitemShut {NoStop}%
\bibitem [{\citenamefont {Suzuki}\ \emph {et~al.}(2017)\citenamefont {Suzuki},
  \citenamefont {Koretsune}, \citenamefont {Ochi},\ and\ \citenamefont
  {Arita}}]{Suzuki:2017}%
  \BibitemOpen
  \bibfield  {author} {\bibinfo {author} {\bibfnamefont {M.-T.}\ \bibnamefont
  {Suzuki}}, \bibinfo {author} {\bibfnamefont {T.}~\bibnamefont {Koretsune}},
  \bibinfo {author} {\bibfnamefont {M.}~\bibnamefont {Ochi}}, \ and\ \bibinfo
  {author} {\bibfnamefont {R.}~\bibnamefont {Arita}},\ }\href {\doibase
  10.1103/PhysRevB.95.094406} {\bibfield  {journal} {\bibinfo  {journal} {Phys.
  Rev. B}\ }\textbf {\bibinfo {volume} {95}},\ \bibinfo {pages} {094406}
  (\bibinfo {year} {2017})}\BibitemShut {NoStop}%
\bibitem [{\citenamefont {Custers}\ \emph {et~al.}(2012)\citenamefont
  {Custers}, \citenamefont {Lorenzer}, \citenamefont {Müller}, \citenamefont
  {Prokofiev}, \citenamefont {Sidorenko}, \citenamefont {Winkler},
  \citenamefont {Strydom}, \citenamefont {Shimura}, \citenamefont {Sakakibara},
  \citenamefont {Yu}, \citenamefont {Si},\ and\ \citenamefont
  {Paschen}}]{Custers2012}%
  \BibitemOpen
  \bibfield  {author} {\bibinfo {author} {\bibfnamefont {J.}~\bibnamefont
  {Custers}}, \bibinfo {author} {\bibfnamefont {K.-A.}\ \bibnamefont
  {Lorenzer}}, \bibinfo {author} {\bibfnamefont {M.}~\bibnamefont {Müller}},
  \bibinfo {author} {\bibfnamefont {A.}~\bibnamefont {Prokofiev}}, \bibinfo
  {author} {\bibfnamefont {A.}~\bibnamefont {Sidorenko}}, \bibinfo {author}
  {\bibfnamefont {H.}~\bibnamefont {Winkler}}, \bibinfo {author} {\bibfnamefont
  {A.~M.}\ \bibnamefont {Strydom}}, \bibinfo {author} {\bibfnamefont
  {Y.}~\bibnamefont {Shimura}}, \bibinfo {author} {\bibfnamefont
  {T.}~\bibnamefont {Sakakibara}}, \bibinfo {author} {\bibfnamefont
  {R.}~\bibnamefont {Yu}}, \bibinfo {author} {\bibfnamefont {Q.}~\bibnamefont
  {Si}}, \ and\ \bibinfo {author} {\bibfnamefont {S.}~\bibnamefont {Paschen}},\
  }\href@noop {} {\bibfield  {journal} {\bibinfo  {journal} {Nature Materials}\
  }\textbf {\bibinfo {volume} {11}},\ \bibinfo {pages} {189} (\bibinfo {year}
  {2012})}\BibitemShut {NoStop}%
\bibitem [{\citenamefont {Martelli}\ \emph {et~al.}(2017)\citenamefont
  {Martelli}, \citenamefont {Cai}, \citenamefont {Nica}, \citenamefont
  {Taupin}, \citenamefont {Prokofiev}, \citenamefont {Liu}, \citenamefont
  {Lai}, \citenamefont {Yu}, \citenamefont {K\"uchler}, \citenamefont
  {Strydom}, \citenamefont {Geiger}, \citenamefont {Haenel}, \citenamefont
  {Larrea}, \citenamefont {Si},\ and\ \citenamefont {Paschen}}]{Martelli2017}%
  \BibitemOpen
  \bibfield  {author} {\bibinfo {author} {\bibfnamefont {V.}~\bibnamefont
  {Martelli}}, \bibinfo {author} {\bibfnamefont {A.}~\bibnamefont {Cai}},
  \bibinfo {author} {\bibfnamefont {E.~M.}\ \bibnamefont {Nica}}, \bibinfo
  {author} {\bibfnamefont {M.}~\bibnamefont {Taupin}}, \bibinfo {author}
  {\bibfnamefont {A.}~\bibnamefont {Prokofiev}}, \bibinfo {author}
  {\bibfnamefont {C.-C.}\ \bibnamefont {Liu}}, \bibinfo {author} {\bibfnamefont
  {H.-H.}\ \bibnamefont {Lai}}, \bibinfo {author} {\bibfnamefont
  {R.}~\bibnamefont {Yu}}, \bibinfo {author} {\bibfnamefont {R.}~\bibnamefont
  {K\"uchler}}, \bibinfo {author} {\bibfnamefont {A.~M.}\ \bibnamefont
  {Strydom}}, \bibinfo {author} {\bibfnamefont {D.}~\bibnamefont {Geiger}},
  \bibinfo {author} {\bibfnamefont {J.}~\bibnamefont {Haenel}}, \bibinfo
  {author} {\bibfnamefont {J.}~\bibnamefont {Larrea}}, \bibinfo {author}
  {\bibfnamefont {Q.}~\bibnamefont {Si}}, \ and\ \bibinfo {author}
  {\bibfnamefont {S.}~\bibnamefont {Paschen}},\ }\href@noop {} {\bibfield
  {journal} {\bibinfo  {journal} {{arXiv:1709.09376}}\ } (\bibinfo {year}
  {2017})}\BibitemShut {NoStop}%
\bibitem [{\citenamefont {Sakai}\ and\ \citenamefont
  {Nakatsuji}(2011)}]{Sakai2011}%
  \BibitemOpen
  \bibfield  {author} {\bibinfo {author} {\bibfnamefont {A.}~\bibnamefont
  {Sakai}}\ and\ \bibinfo {author} {\bibfnamefont {S.}~\bibnamefont
  {Nakatsuji}},\ }\href {\doibase 10.1143/JPSJ.80.063701} {\bibfield  {journal}
  {\bibinfo  {journal} {Journal of the Physical Society of Japan}\ }\textbf
  {\bibinfo {volume} {80}},\ \bibinfo {pages} {063701} (\bibinfo {year}
  {2011})}\BibitemShut {NoStop}%
\bibitem [{\citenamefont {Lee}\ \emph {et~al.}(2018)\citenamefont {Lee},
  \citenamefont {Trebst}, \citenamefont {Kim},\ and\ \citenamefont
  {Paramekanti}}]{Lee2018}%
  \BibitemOpen
  \bibfield  {author} {\bibinfo {author} {\bibfnamefont {S.}~\bibnamefont
  {Lee}}, \bibinfo {author} {\bibfnamefont {S.}~\bibnamefont {Trebst}},
  \bibinfo {author} {\bibfnamefont {Y.~B.}\ \bibnamefont {Kim}}, \ and\
  \bibinfo {author} {\bibfnamefont {A.}~\bibnamefont {Paramekanti}},\
  }\href@noop {} {\bibfield  {journal} {\bibinfo  {journal} {arXiv:1806.02842}\
  } (\bibinfo {year} {2018})}\BibitemShut {NoStop}%
\bibitem [{\citenamefont {Bauer}\ \emph {et~al.}(2002)\citenamefont {Bauer},
  \citenamefont {Frederick}, \citenamefont {Ho}, \citenamefont {Zapf},\ and\
  \citenamefont {Maple}}]{Bauer2002}%
  \BibitemOpen
  \bibfield  {author} {\bibinfo {author} {\bibfnamefont {E.~D.}\ \bibnamefont
  {Bauer}}, \bibinfo {author} {\bibfnamefont {N.~A.}\ \bibnamefont
  {Frederick}}, \bibinfo {author} {\bibfnamefont {P.-C.}\ \bibnamefont {Ho}},
  \bibinfo {author} {\bibfnamefont {V.~S.}\ \bibnamefont {Zapf}}, \ and\
  \bibinfo {author} {\bibfnamefont {M.~B.}\ \bibnamefont {Maple}},\ }\href
  {\doibase 10.1103/PhysRevB.65.100506} {\bibfield  {journal} {\bibinfo
  {journal} {Phys. Rev. B}\ }\textbf {\bibinfo {volume} {65}},\ \bibinfo
  {pages} {100506} (\bibinfo {year} {2002})}\BibitemShut {NoStop}%
\bibitem [{\citenamefont {McCollam}\ \emph {et~al.}(2013)\citenamefont
  {McCollam}, \citenamefont {Andraka},\ and\ \citenamefont
  {Julian}}]{McCollam2013}%
  \BibitemOpen
  \bibfield  {author} {\bibinfo {author} {\bibfnamefont {A.}~\bibnamefont
  {McCollam}}, \bibinfo {author} {\bibfnamefont {B.}~\bibnamefont {Andraka}}, \
  and\ \bibinfo {author} {\bibfnamefont {S.~R.}\ \bibnamefont {Julian}},\
  }\href {\doibase 10.1103/PhysRevB.88.075102} {\bibfield  {journal} {\bibinfo
  {journal} {Phys. Rev. B}\ }\textbf {\bibinfo {volume} {88}},\ \bibinfo
  {pages} {075102} (\bibinfo {year} {2013})}\BibitemShut {NoStop}%
\bibitem [{\citenamefont {Si}\ and\ \citenamefont
  {Steglich}(2010)}]{SiSteglich2010}%
  \BibitemOpen
  \bibfield  {author} {\bibinfo {author} {\bibfnamefont {Q.}~\bibnamefont
  {Si}}\ and\ \bibinfo {author} {\bibfnamefont {F.}~\bibnamefont {Steglich}},\
  }\href {\doibase 10.1126/science.1191195} {\bibfield  {journal} {\bibinfo
  {journal} {Science}\ }\textbf {\bibinfo {volume} {329}},\ \bibinfo {pages}
  {1161} (\bibinfo {year} {2010})}\BibitemShut {NoStop}%
\bibitem [{\citenamefont {Coleman}\ and\ \citenamefont
  {Schofield}(2005)}]{ColemanSchofield2005}%
  \BibitemOpen
  \bibfield  {author} {\bibinfo {author} {\bibfnamefont {P.}~\bibnamefont
  {Coleman}}\ and\ \bibinfo {author} {\bibfnamefont {A.~J.}\ \bibnamefont
  {Schofield}},\ }\href {\doibase 10.1038/nature03279} {\bibfield  {journal}
  {\bibinfo  {journal} {Nature}\ }\textbf {\bibinfo {volume} {433}},\ \bibinfo
  {pages} {226} (\bibinfo {year} {2005})}\BibitemShut {NoStop}%
\bibitem [{\citenamefont {Si}\ \emph {et~al.}(2001)\citenamefont {Si},
  \citenamefont {Rabello}, \citenamefont {Ingersent},\ and\ \citenamefont
  {Smith}}]{Si-Nature}%
  \BibitemOpen
  \bibfield  {author} {\bibinfo {author} {\bibfnamefont {Q.}~\bibnamefont
  {Si}}, \bibinfo {author} {\bibfnamefont {S.}~\bibnamefont {Rabello}},
  \bibinfo {author} {\bibfnamefont {K.}~\bibnamefont {Ingersent}}, \ and\
  \bibinfo {author} {\bibfnamefont {J.~L.}\ \bibnamefont {Smith}},\ }\href
  {\doibase 10.1038/35101507} {\bibfield  {journal} {\bibinfo  {journal}
  {Nature}\ }\textbf {\bibinfo {volume} {413}},\ \bibinfo {pages} {804}
  (\bibinfo {year} {2001})}\BibitemShut {NoStop}%
\bibitem [{\citenamefont {Coleman}\ \emph {et~al.}(2001)\citenamefont
  {Coleman}, \citenamefont {P\'{e}pin}, \citenamefont {Si},\ and\ \citenamefont
  {Ramazashvili}}]{Colemanetal}%
  \BibitemOpen
  \bibfield  {author} {\bibinfo {author} {\bibfnamefont {P.}~\bibnamefont
  {Coleman}}, \bibinfo {author} {\bibfnamefont {C.}~\bibnamefont {P\'{e}pin}},
  \bibinfo {author} {\bibfnamefont {Q.}~\bibnamefont {Si}}, \ and\ \bibinfo
  {author} {\bibfnamefont {R.}~\bibnamefont {Ramazashvili}},\ }\href
  {http://stacks.iop.org/0953-8984/13/i=35/a=202} {\bibfield  {journal}
  {\bibinfo  {journal} {Journal of Physics: Condensed Matter}\ }\textbf
  {\bibinfo {volume} {13}},\ \bibinfo {pages} {R723} (\bibinfo {year}
  {2001})}\BibitemShut {NoStop}%
\bibitem [{\citenamefont {Yamamoto}\ and\ \citenamefont
  {Si}(2007)}]{Yamamoto2007}%
  \BibitemOpen
  \bibfield  {author} {\bibinfo {author} {\bibfnamefont {S.~J.}\ \bibnamefont
  {Yamamoto}}\ and\ \bibinfo {author} {\bibfnamefont {Q.}~\bibnamefont {Si}},\
  }\href {\doibase 10.1103/PhysRevLett.99.016401} {\bibfield  {journal}
  {\bibinfo  {journal} {Phys. Rev. Lett.}\ }\textbf {\bibinfo {volume} {99}},\
  \bibinfo {pages} {016401} (\bibinfo {year} {2007})}\BibitemShut {NoStop}%
\bibitem [{\citenamefont {Portnichenko}\ \emph {et~al.}(2016)\citenamefont
  {Portnichenko}, \citenamefont {Paschen}, \citenamefont {Prokofiev},
  \citenamefont {Vojta}, \citenamefont {Cameron}, \citenamefont {Mignot},
  \citenamefont {Ivanov},\ and\ \citenamefont {Inosov}}]{Por16.1}%
  \BibitemOpen
  \bibfield  {author} {\bibinfo {author} {\bibfnamefont {P.~Y.}\ \bibnamefont
  {Portnichenko}}, \bibinfo {author} {\bibfnamefont {S.}~\bibnamefont
  {Paschen}}, \bibinfo {author} {\bibfnamefont {A.}~\bibnamefont {Prokofiev}},
  \bibinfo {author} {\bibfnamefont {M.}~\bibnamefont {Vojta}}, \bibinfo
  {author} {\bibfnamefont {A.~S.}\ \bibnamefont {Cameron}}, \bibinfo {author}
  {\bibfnamefont {J.-M.}\ \bibnamefont {Mignot}}, \bibinfo {author}
  {\bibfnamefont {A.}~\bibnamefont {Ivanov}}, \ and\ \bibinfo {author}
  {\bibfnamefont {D.~S.}\ \bibnamefont {Inosov}},\ }\href {\doibase
  10.1103/PhysRevB.94.245132} {\bibfield  {journal} {\bibinfo  {journal}
  {{Phys.\ Rev.\ B}}\ }\textbf {\bibinfo {volume} {94}},\ \bibinfo {pages}
  {245132} (\bibinfo {year} {2016})}\BibitemShut {NoStop}%
\bibitem [{\citenamefont {Yamamoto}\ and\ \citenamefont
  {Si}(2010)}]{Yamamoto2010}%
  \BibitemOpen
  \bibfield  {author} {\bibinfo {author} {\bibfnamefont {S.~J.}\ \bibnamefont
  {Yamamoto}}\ and\ \bibinfo {author} {\bibfnamefont {Q.}~\bibnamefont {Si}},\
  }\href {\doibase 10.1007/s10909-010-0221-4} {\bibfield  {journal} {\bibinfo
  {journal} {Journal of Low Temperature Physics}\ }\textbf {\bibinfo {volume}
  {161}},\ \bibinfo {pages} {233} (\bibinfo {year} {2010})}\BibitemShut
  {NoStop}%
\bibitem [{\citenamefont {Hu}\ \emph {et~al.}(2017)\citenamefont {Hu},
  \citenamefont {Gong}, \citenamefont {Lai}, \citenamefont {Hu}, \citenamefont
  {Si},\ and\ \citenamefont {Nevidomskyy}}]{Wenjun_NSL2017}%
  \BibitemOpen
  \bibfield  {author} {\bibinfo {author} {\bibfnamefont {W.-J.}\ \bibnamefont
  {Hu}}, \bibinfo {author} {\bibfnamefont {S.-S.}\ \bibnamefont {Gong}},
  \bibinfo {author} {\bibfnamefont {H.-H.}\ \bibnamefont {Lai}}, \bibinfo
  {author} {\bibfnamefont {H.}~\bibnamefont {Hu}}, \bibinfo {author}
  {\bibfnamefont {Q.}~\bibnamefont {Si}}, \ and\ \bibinfo {author}
  {\bibfnamefont {A.~H.}\ \bibnamefont {Nevidomskyy}},\ }\href@noop {}
  {\bibfield  {journal} {\bibinfo  {journal} {arXiv:1711.06523}\ } (\bibinfo
  {year} {2017})}\BibitemShut {NoStop}%
\bibitem [{sup()}]{supp}%
  \BibitemOpen
  \href@noop {} {}\bibinfo {note} {{See Supplemental Material for more details
  of derivations[url]}}\BibitemShut {NoStop}%
\bibitem [{\citenamefont {Paschen}\ and\ \citenamefont {{Larrea
  J.}}(2014)}]{Pas14.1}%
  \BibitemOpen
  \bibfield  {author} {\bibinfo {author} {\bibfnamefont {S.}~\bibnamefont
  {Paschen}}\ and\ \bibinfo {author} {\bibfnamefont {J.}~\bibnamefont {{Larrea
  J.}}},\ }\href@noop {} {\bibfield  {journal} {\bibinfo  {journal} {{J.\
  Phys.\ Soc.\ Jpn.}}\ }\textbf {\bibinfo {volume} {83}},\ \bibinfo {pages}
  {061004} (\bibinfo {year} {2014})}\BibitemShut {NoStop}%
\bibitem [{\citenamefont {Nakao}\ \emph {et~al.}(2001)\citenamefont {Nakao},
  \citenamefont {Magishi}, \citenamefont {Wakabayashi}, \citenamefont
  {Murakami}, \citenamefont {Koyama}, \citenamefont {Hirota}, \citenamefont
  {Endoh},\ and\ \citenamefont {Kunii}}]{Nak01.1}%
  \BibitemOpen
  \bibfield  {author} {\bibinfo {author} {\bibfnamefont {H.}~\bibnamefont
  {Nakao}}, \bibinfo {author} {\bibfnamefont {K.}~\bibnamefont {Magishi}},
  \bibinfo {author} {\bibfnamefont {Y.}~\bibnamefont {Wakabayashi}}, \bibinfo
  {author} {\bibfnamefont {Y.}~\bibnamefont {Murakami}}, \bibinfo {author}
  {\bibfnamefont {K.}~\bibnamefont {Koyama}}, \bibinfo {author} {\bibfnamefont
  {K.}~\bibnamefont {Hirota}}, \bibinfo {author} {\bibfnamefont
  {Y.}~\bibnamefont {Endoh}}, \ and\ \bibinfo {author} {\bibfnamefont
  {S.}~\bibnamefont {Kunii}},\ }\href {\doibase 10.1143/JPSJ.70.1857}
  {\bibfield  {journal} {\bibinfo  {journal} {{J.\ Phys.\ Soc.\ Jpn.}}\
  }\textbf {\bibinfo {volume} {70}},\ \bibinfo {pages} {1857} (\bibinfo {year}
  {2001})}\BibitemShut {NoStop}%
\bibitem [{\citenamefont {Kawarasaki}\ \emph {et~al.}(2011)\citenamefont
  {Kawarasaki}, \citenamefont {Matsumura}, \citenamefont {Sera},\ and\
  \citenamefont {Ochiai}}]{Kaw11.1}%
  \BibitemOpen
  \bibfield  {author} {\bibinfo {author} {\bibfnamefont {Y.}~\bibnamefont
  {Kawarasaki}}, \bibinfo {author} {\bibfnamefont {T.}~\bibnamefont
  {Matsumura}}, \bibinfo {author} {\bibfnamefont {M.}~\bibnamefont {Sera}}, \
  and\ \bibinfo {author} {\bibfnamefont {A.}~\bibnamefont {Ochiai}},\
  }\href@noop {} {\bibfield  {journal} {\bibinfo  {journal} {{J.\ Phys.\ Soc.\
  Jpn.}}\ }\textbf {\bibinfo {volume} {80}},\ \bibinfo {pages} {023713}
  (\bibinfo {year} {2011})}\BibitemShut {NoStop}%
\bibitem [{\citenamefont {Tanida}\ \emph {et~al.}(2018)\citenamefont {Tanida},
  \citenamefont {Muro},\ and\ \citenamefont {Matsumura}}]{Tan18.1}%
  \BibitemOpen
  \bibfield  {author} {\bibinfo {author} {\bibfnamefont {H.}~\bibnamefont
  {Tanida}}, \bibinfo {author} {\bibfnamefont {Y.}~\bibnamefont {Muro}}, \ and\
  \bibinfo {author} {\bibfnamefont {T.}~\bibnamefont {Matsumura}},\ }\href
  {\doibase 10.7566/JPSJ.87.023705} {\bibfield  {journal} {\bibinfo  {journal}
  {{J.\ Phys.\ Soc.\ Jpn.}}\ }\textbf {\bibinfo {volume} {87}},\ \bibinfo
  {pages} {023705} (\bibinfo {year} {2018})}\BibitemShut {NoStop}%
\end{thebibliography}%


\begin{thebibliography}{6}%
\makeatletter
\providecommand \@ifxundefined [1]{%
 \@ifx{#1\undefined}
}%
\providecommand \@ifnum [1]{%
 \ifnum #1\expandafter \@firstoftwo
 \else \expandafter \@secondoftwo
 \fi
}%
\providecommand \@ifx [1]{%
 \ifx #1\expandafter \@firstoftwo
 \else \expandafter \@secondoftwo
 \fi
}%
\providecommand \natexlab [1]{#1}%
\providecommand \enquote  [1]{``#1''}%
\providecommand \bibnamefont  [1]{#1}%
\providecommand \bibfnamefont [1]{#1}%
\providecommand \citenamefont [1]{#1}%
\providecommand \href@noop [0]{\@secondoftwo}%
\providecommand \href [0]{\begingroup \@sanitize@url \@href}%
\providecommand \@href[1]{\@@startlink{#1}\@@href}%
\providecommand \@@href[1]{\endgroup#1\@@endlink}%
\providecommand \@sanitize@url [0]{\catcode `\\12\catcode `\$12\catcode
  `\&12\catcode `\#12\catcode `\^12\catcode `\_12\catcode `\%12\relax}%
\providecommand \@@startlink[1]{}%
\providecommand \@@endlink[0]{}%
\providecommand \url  [0]{\begingroup\@sanitize@url \@url }%
\providecommand \@url [1]{\endgroup\@href {#1}{\urlprefix }}%
\providecommand \urlprefix  [0]{URL }%
\providecommand \Eprint [0]{\href }%
\providecommand \doibase [0]{http://dx.doi.org/}%
\providecommand \selectlanguage [0]{\@gobble}%
\providecommand \bibinfo  [0]{\@secondoftwo}%
\providecommand \bibfield  [0]{\@secondoftwo}%
\providecommand \translation [1]{[#1]}%
\providecommand \BibitemOpen [0]{}%
\providecommand \bibitemStop [0]{}%
\providecommand \bibitemNoStop [0]{.\EOS\space}%
\providecommand \EOS [0]{\spacefactor3000\relax}%
\providecommand \BibitemShut  [1]{\csname bibitem#1\endcsname}%
\let\auto@bib@innerbib\@empty
\bibitem [{\citenamefont {White}(1992)}]{white1992}%
  \BibitemOpen
  \bibfield  {author} {\bibinfo {author} {\bibfnamefont {S.~R.}\ \bibnamefont
  {White}},\ }\href {\doibase 10.1103/PhysRevLett.69.2863} {\bibfield
  {journal} {\bibinfo  {journal} {Phys. Rev. Lett.}\ }\textbf {\bibinfo
  {volume} {69}},\ \bibinfo {pages} {2863} (\bibinfo {year}
  {1992})}\BibitemShut {NoStop}%
\bibitem [{\citenamefont {McCulloch}\ and\ \citenamefont
  {Gul{\'a}csi}(2002)}]{mcculloch2002}%
  \BibitemOpen
  \bibfield  {author} {\bibinfo {author} {\bibfnamefont {I.}~\bibnamefont
  {McCulloch}}\ and\ \bibinfo {author} {\bibfnamefont {M.}~\bibnamefont
  {Gul{\'a}csi}},\ }\href {http://iopscience.iop.org/0295-5075/57/6/852}
  {\bibfield  {journal} {\bibinfo  {journal} {Europhysics Letters}\ }\textbf
  {\bibinfo {volume} {57}},\ \bibinfo {pages} {852} (\bibinfo {year}
  {2002})}\BibitemShut {NoStop}%
\bibitem [{\citenamefont {Blume}\ and\ \citenamefont
  {Hsieh}(1969)}]{blume1969}%
  \BibitemOpen
  \bibfield  {author} {\bibinfo {author} {\bibfnamefont {M.}~\bibnamefont
  {Blume}}\ and\ \bibinfo {author} {\bibfnamefont {Y.~Y.}\ \bibnamefont
  {Hsieh}},\ }\href {\doibase http://dx.doi.org/10.1063/1.1657616} {\bibfield
  {journal} {\bibinfo  {journal} {Journal of Applied Physics}\ }\textbf
  {\bibinfo {volume} {40}},\ \bibinfo {pages} {1249} (\bibinfo {year}
  {1969})}\BibitemShut {NoStop}%
\bibitem [{\citenamefont {{Hu}}\ \emph {et~al.}(2016)\citenamefont {{Hu}},
  \citenamefont {{Lai}}, \citenamefont {{Gong}}, \citenamefont {{Yu}},
  \citenamefont {{Nevidomskyy}},\ and\ \citenamefont {{Si}}}]{hufese}%
  \BibitemOpen
  \bibfield  {author} {\bibinfo {author} {\bibfnamefont {W.-J.}\ \bibnamefont
  {{Hu}}}, \bibinfo {author} {\bibfnamefont {H.-H.}\ \bibnamefont {{Lai}}},
  \bibinfo {author} {\bibfnamefont {S.-S.}\ \bibnamefont {{Gong}}}, \bibinfo
  {author} {\bibfnamefont {R.}~\bibnamefont {{Yu}}}, \bibinfo {author}
  {\bibfnamefont {A.~H.}\ \bibnamefont {{Nevidomskyy}}}, \ and\ \bibinfo
  {author} {\bibfnamefont {Q.}~\bibnamefont {{Si}}},\ }\href@noop {} {\bibfield
   {journal} {\bibinfo  {journal} {ArXiv e-prints}\ } (\bibinfo {year}
  {2016})},\ \Eprint {http://arxiv.org/abs/1606.01235} {arXiv:1606.01235
  [cond-mat.str-el]} \BibitemShut {NoStop}%
\bibitem [{\citenamefont {Smerald}\ and\ \citenamefont
  {Shannon}(2013)}]{Smerald2013}%
  \BibitemOpen
  \bibfield  {author} {\bibinfo {author} {\bibfnamefont {A.}~\bibnamefont
  {Smerald}}\ and\ \bibinfo {author} {\bibfnamefont {N.}~\bibnamefont
  {Shannon}},\ }\href {\doibase 10.1103/PhysRevB.88.184430} {\bibfield
  {journal} {\bibinfo  {journal} {Phys. Rev. B}\ }\textbf {\bibinfo {volume}
  {88}},\ \bibinfo {pages} {184430} (\bibinfo {year} {2013})}\BibitemShut
  {NoStop}%
\bibitem [{\citenamefont {Yamamoto}\ and\ \citenamefont
  {Si}(2010)}]{Yamamoto2010}%
  \BibitemOpen
  \bibfield  {author} {\bibinfo {author} {\bibfnamefont {S.~J.}\ \bibnamefont
  {Yamamoto}}\ and\ \bibinfo {author} {\bibfnamefont {Q.}~\bibnamefont {Si}},\
  }\href {\doibase 10.1007/s10909-010-0221-4} {\bibfield  {journal} {\bibinfo
  {journal} {Journal of Low Temperature Physics}\ }\textbf {\bibinfo {volume}
  {161}},\ \bibinfo {pages} {233} (\bibinfo {year} {2010})}\BibitemShut
  {NoStop}%
\end{thebibliography}%
\end{document}


\title{Supplemental Material for Kondo destruction in multipolar order and implications for heavy-fermion quantum criticality} 
\author{Hsin-Hua Lai}
\affiliation{Department of Physics and Astronomy, Rice University, Houston, Texas 77005, USA}
\author{Emilian M. Nica}
\affiliation{Department of Physics and Astronomy and Quantum Materials Institute,
University of British Columbia, Vancouver, B.C., V6T 1Z1, Canada}
\author{Wen-Jun Hu}
\affiliation{Department of Physics and Astronomy, Rice University, Houston, Texas 77005, USA}
\author{Shou-Shu Gong}
\affiliation{Department of Physics and International Research Institute of Multidisciplinary Science, Beihang University, Beijing 100191, China}
\author{Qimiao Si}
\affiliation{Department of Physics and Astronomy, Rice University, Houston, Texas 77005, USA}

\date{\today}
\begin{abstract}
In this supplemental material, we give more details on the DMRG calculations and the derivations for parts of the main text.
\end{abstract}

\maketitle

\section{DMRG results for the spin structure factor}
\begin{figure}
   \includegraphics[width= 3 in]{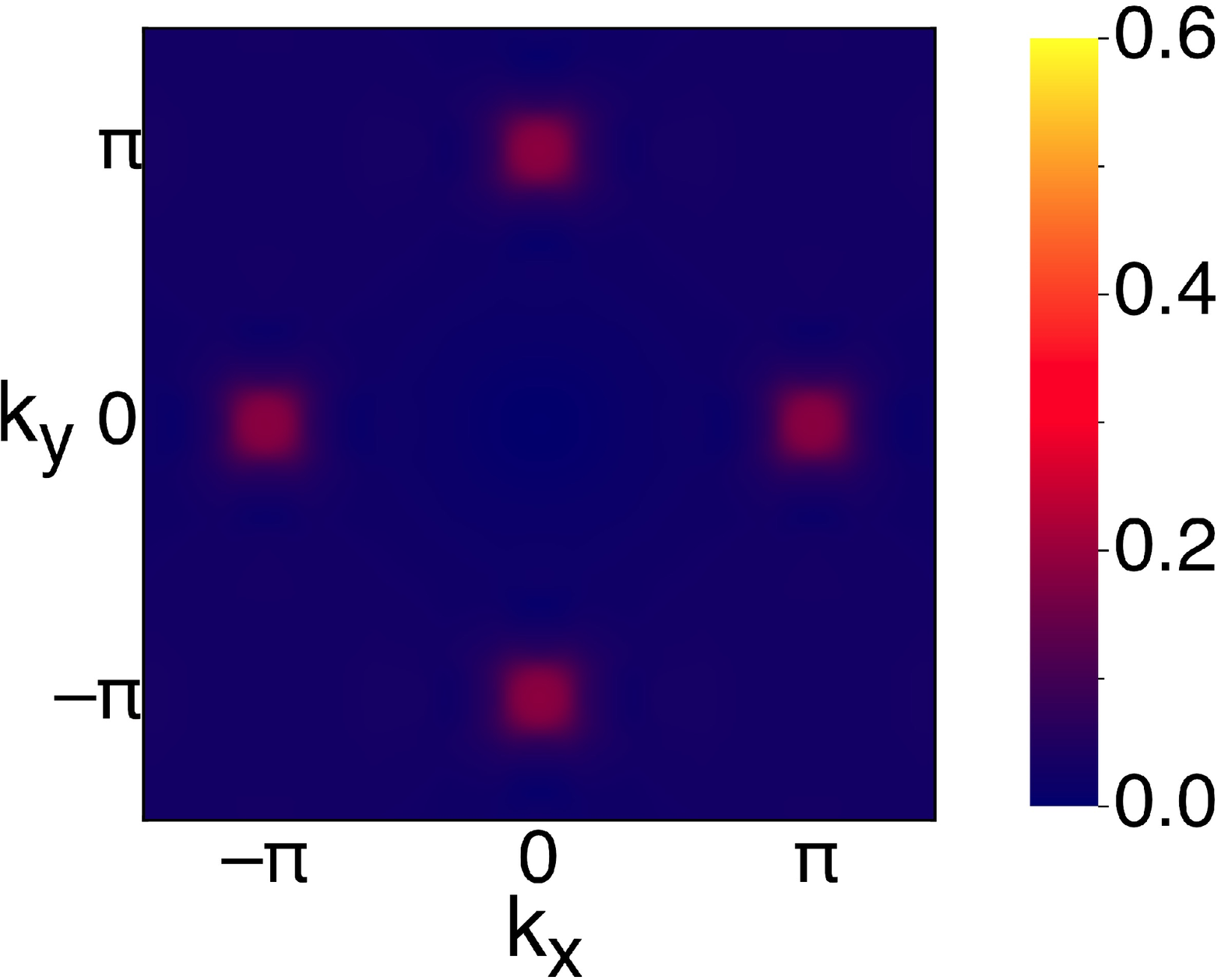} 
   \caption{(Color online) The spin structure factor $m^2_{\bm{S}}$ by DMRG with $J_1=1,K_1 = 1.2,K_2=-0.3$ on the $L_y=8$ cylinder.}
   \label{msq}
\end{figure}

The DMRG~\cite{white1992, mcculloch2002} simulations have been performed on the geometry of a rectangular cylinder, with 
periodic boundary conditions in the $y$ direction and open boundaries in the $x$ direction. 
We study the system with $L_y$ up to $8$. By keeping up to $4000$ SU(2) DMRG states, 
our calculations have the largest truncation errors around $10^{-5}$ to show high accuracy. We calculate the spin-spin 
(${\bf S}_{i}\cdot {\bf S}_{j}$) and quadrupolar-quadrupolar (${\bf Q}_{i}\cdot {\bf Q}_{j}$) correlation functions, 
where ${\bf Q}_{i}$ is the quadrupolar operator~\cite{blume1969} and the quadrupolar term can be reexpressed as ${\bf Q}_{i}\cdot {\bf Q}_{j}=2({\bf S}_{i}\cdot {\bf S}_{j})^2+{\bf S}_{i}\cdot {\bf S}_{j}-8/3$. We perform Fourier transformation for the correlation functions 
to obtain the spin and quadrupolar structure factors as 
\begin{equation}
m^{2}_{\bm{S}}({\bf q}) = \frac{1}{N_s^2}\sum_{i,j} \langle {\bf S}_{i}\cdot {\bf S}_{j} \rangle e^{i{\bf q}\cdot({\bf r}_i-{\bf r}_j)}
\end{equation}
and 
\begin{equation}
m^{2}_{\bm{S}}({\bf q}) = \frac{1}{N_s^2}\sum_{i,j} \langle {\bf Q}_{i}\cdot {\bf Q}_{j} \rangle e^{i{\bf q}\cdot({\bf r}_i - {\bf r}_j)},
\end{equation}
where the sites $i, j$ are chosen over the middle $N_s = L_y \times L_y$ sites in order to avoid the effects of open edge~\cite{hufese}.
Fig.~\ref{msq} shows the spin structure factor $m^2_{\bm{S}}$ with $J_1=1,K_1 = 1.2,K_2=-0.3$ on the $L_y=8$ cylinder, which has weak peaks at momenta $(\pi,0)$ and $(0,\pi)$. The quadrupolar structure factor $m^2_{\bm{Q}}$ has been shown in the main text. 

\section{3-flavor electron representations of $S=1$ spin and quadrupole}
For conduction electrons, we consider 3-flavor electrons with flavor index $\alpha = x, y, z$. The three flavor of fermions can be transformed to each other by SU(3) symmetry. For each lattice site, the total electron density is one, which means that each flavor of electron is 1/3-filled
\begin{eqnarray}
\left\la \psi^\dagger_{i \alpha}\psi_{i \alpha}\right\ra = \frac{1}{3}
\end{eqnarray}
The spin and quadrupole operators can be written in the 3-flavor electrons as,
\begin{eqnarray}
&& s^\alpha_{c,i} = -i\epsilon_{\alpha \beta \gamma} \psi_{i \beta}^\dagger \psi_{i \gamma},\\
&& q^{x^2 - y^2}_{c,i} = -\psi^\dagger_{ix}\psi_{ix} +\psi^\dagger_{iy}\psi_{iy}, \\
&& q^{3z^2 - r^2}_{c,i} = (\psi_{ix}^\dagger \psi_{ix} + \psi_{iy}^\dagger \psi_{iy} - 2 \psi^\dagger_{iz}\psi_{iz})/\sqrt{3}, \\
&& q^{xy}_{c,i} = - \psi^\dagger_{ix}\psi_{iy} - \psi^\dagger_{i y}\psi_{i x},\\
&& q^{yz}_{c,i} = - \psi^\dagger_{iy}\psi_{iz} - \psi^\dagger_{i z}\psi_{i y},\\
&& q^{zx}_{c, i} = - \psi^\dagger_{iz}\psi_{ix} - \psi^\dagger_{i x}\psi_{i z}.
\end{eqnarray}
The spin and quadrupolar operators can also be transformed to each other under the SU(3) rotations.

\section{Continuum theory at SU(3) point}
Let's focus first on the SU(3) point, where $J_n = K_n$. We consider the time-reversal invariance basis
\begin{equation}
\begin{array}{ccc}
|x\ra = i \frac{| 1 \ra - |\bar{1}\ra}{\sqrt{2}}, & |y\ra = \frac{| 1\ra + | \bar{1} \ra}{\sqrt{2}}, & |z \ra = -i | 0 \ra,
\end{array}
\end{equation}
where $|S^z = 1\ra \equiv |1\ra$ and etc. A general wave function at a site $j$ can be written in the form
\begin{eqnarray}
|{\bf d}_j\ra = d^x_j |x\ra + d^y_j |y \ra + d^z_j | z\ra,
\end{eqnarray} 
where we can introduce the vector notation as
\begin{eqnarray}
{\bf d}_j = (d^x_j,~d^y_j,~d^z_j)
\end{eqnarray}
is a 3 vector of complex numbers. Separating the real and imaginary parts of ${\bf d}_j$ gives
\begin{eqnarray}
{\bf d}_j = {\bf u}_j + i {\bf v}_j.
\end{eqnarray}
Requiring the wave functions to be normalizd gives the constraints,
\begin{eqnarray}
{\bf d}_j \cdot \bar{\bf d}_j = 1 \rightarrow {\bf u}^2_j + {\bf v}^2_j =1.
\end{eqnarray}
The overall phase can be fixed by requiring
\begin{eqnarray}
{\bf d}_j^2 = \bar{\bf d}_j^2 \rightarrow {\bf u}_j \cdot {\bf v}_j = 0.
\end{eqnarray}
Within the spin-coherent state framework, we can obtain
\begin{eqnarray}
{\bf S}_j = 2 {\bf u}_j \times {\bf v}_j,
\end{eqnarray}
and in terms of the components of the ${\bf d}$ we can obtain
\begin{eqnarray}
&& S^\alpha = - i \epsilon^{\alpha \beta \gamma} \bar{d}^\beta d^\gamma, \\
&& Q^{x^2 - y^2} = - |d^x|^2 + |d^y|^2,\\
&& Q^{3z^2 - r^2} = \frac{1}{\sqrt{3}}\left[ |d^x|^2 + |d^y|^2 -2 |d^z|^2 \right], \\
&& Q^{xy} = - \bar{d}^x d^y - \bar{d}^y d^x,\\
&& Q^{yz} = - \bar{d}^y d^z - \bar{d}^z d^y, \\
&& Q^{zx} = - \bar{d}^z d^x - \bar{d}^x d^z.
\end{eqnarray} 
The Hamiltonian can be re-expressed as
\begin{eqnarray}
\nonumber H_S = && \sum_{\la i j \ra} \left[ J_1 |{\bf d}_i \cdot \bar{\bf d}_j|^2 + (K_1 - J_1) | {\bf d}_i \cdot {\bf d}_j |^2 + K_1 \right] + \sum_{\la\la i j \ra\ra} \left[ J_2 |{\bf d}_i \cdot \bar{\bf d}_j|^2 +(K_2 - J_2) | {\bf d}_i \cdot {\bf d}_j |^2 + K_2 \right]\\
\nonumber \equiv && \sum_{\la i j \ra} \left[ J_1 |{\bf d}_i \cdot \bar{\bf d}_j|^2 + J_1' | {\bf d}_i \cdot {\bf d}_j |^2 + K_1 \right] + \sum_{\la\la i j \ra\ra} \left[ J_2 |{\bf d}_i \cdot \bar{\bf d}_j|^2 + J_2'| {\bf d}_i \cdot {\bf d}_j |^2 + K_2 \right],
\end{eqnarray}
where we defined $K_n - J_n \equiv J_n'$ which break the SU(3) symmetry and from now on we will ignore the constants. At SU(3) point, we can see that that Hamiltonian becomes
\begin{eqnarray}\label{Eq:SU(3)}
H^{SU(3)}_{S} = \sum_{\la i j \ra} J_1 |{\bf d}_i \cdot \bar{\bf d}_j|^2 + \sum_{\la \la i j \ra \ra} J_2 | {\bf d}_i \cdot \bar{\bf d}_j |^2 .
\end{eqnarray} 

\begin{figure}[t] 
   \includegraphics[width=\columnwidth]{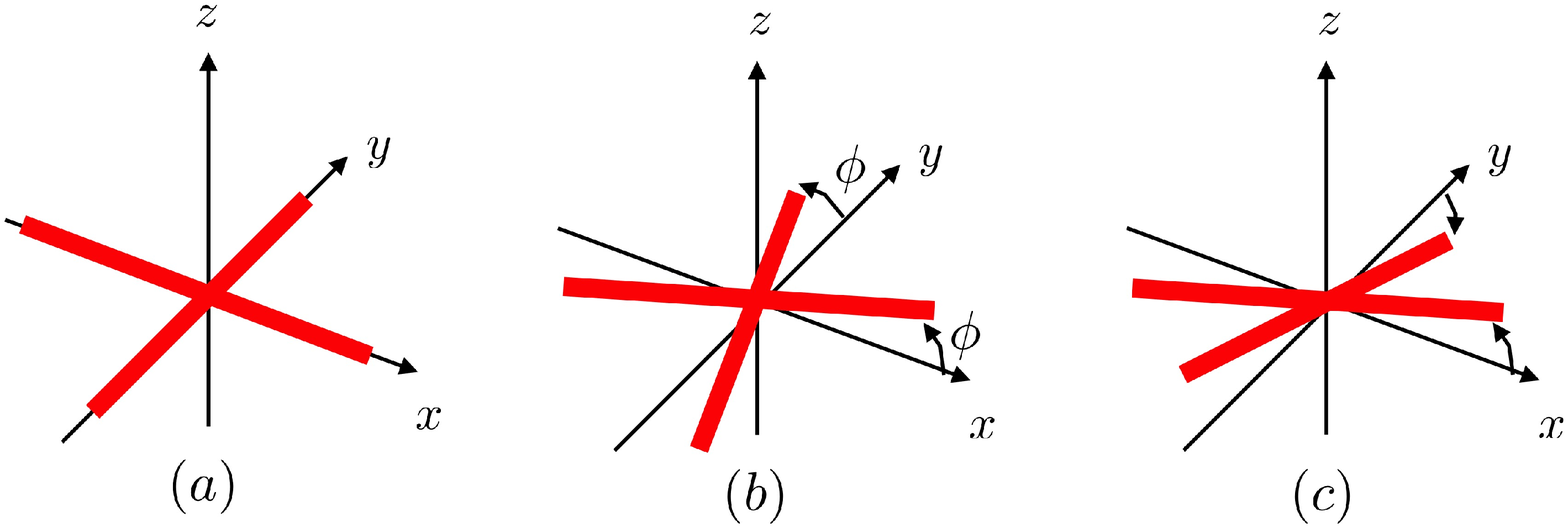} 
   \caption{Illustration of the effects of rotations in the complex director configurations space for $(\pi,\pi)$-AFQ. (a) Illustration of the real component of the directors (red cylinders). (b) Effects of the global rotation around $\hat{z}$-axis, which preserve the angle between the directors (c) Illustration of the rotation using one of the canting fields, $\mu_1$, which changes the angle between the directors that increases the energy.}
   \label{Fig:Rotation}
\end{figure}

For obtaining the non-linear sigma model (NLsM) description, we can follow the approaches detailed by A. Smerald\etal~\cite{Smerald2013} One choice for the ground state of such a $(\pi,\pi)$-AFQ is
\begin{equation}
\begin{array}{cc}
{\bf d}^{gs}_A = \begin{pmatrix}
1\\
0\\
0
\end{pmatrix}, & {\bf d}^{gs}_B = \begin{pmatrix}
0 \\
1 \\
 0
\end{pmatrix}.
\end{array}
\end{equation}
The Hamiltonian, Eq.~(\ref{Eq:SU(3)}), is invariant under the global rotation ${\bf d} \rightarrow {\bf U}{\bf d}$, provided that ${\bf U}^{-1} = {\bf U}^\dagger$. In general, there should be 6 distinct generators for SU(3), however, for the present $(\pi,\pi)$-AFQ phase with only two directors, only 4 out of 6 are needed for a complete description of the global rotations which maintain the energy of the ground state. Fig.~\ref{Fig:Rotation} illustrate the effects of all possible rotations, some of which preserve the energy while some increase the energy. Let's first study the global ratations.

The natural choices of the matrices at SU(3) are Gell-Mann matrices, which can be used to describe the rotation of the ground state configuration. Ignoring the diagonal Gell-Mann matrices, we are left with 6 matrices. However, only 4 are needed for a complete description of the global rotations for the $(\pi,\pi)$-AFQ with only two mutually orthogonal directors.  The 4 Gell-Mann matrices I choose are,
\begin{equation}
\begin{array}{cccc}
\lambda_1 = \begin{pmatrix}
0 & -i & 0 \\
i & 0 & 0 \\
0 & 0 & 0 
\end{pmatrix}
& \lambda_2 = \begin{pmatrix}
0 & 0 & i \\
0 & 0 & 0 \\
-i & 0 & 0
\end{pmatrix} &~~~~
\lambda_3 = \begin{pmatrix}
0 & 0 & 0 \\
0 & 0 & -i \\
0 & i & 0 
\end{pmatrix}
& \lambda_4 = \begin{pmatrix}
0 & 1 & 0 \\
1 & 0 & 0 \\
0 & 0 & 0
\end{pmatrix}.
\end{array}
\end{equation}
The global rotation in complex space can be written as
\begin{eqnarray}
{\bf U}(\phi) \equiv e^{i\sum_{j =1}^4 \lambda_j \phi_j}.
\end{eqnarray}
Fig.~\ref{Fig:Rotation} (b) gives an illustration of the global rotation around real $\hat{z}$-axis using ${\bf U}(\phi_1, 0 ,0 ,0)$.

On the other hand, we also need to consider the rotations involving canting of the directors ground state configurations, which increases the energy. The canting fields I choose are
\begin{equation}
\begin{array}{cccc}
\mu_1 = \begin{pmatrix}
0 & -i & 0\\
-i & 0 & 0 \\
0 & 0 &0 
\end{pmatrix}
& \mu_2 = \begin{pmatrix}
0 & 0 &-1 \\
0 & 0 & 0 \\
1 & 0 & 0
\end{pmatrix} &
\mu_3 = \begin{pmatrix}
0 & 0 & 0 \\
0 & 0 & 1 \\
0 & -1 & 0
\end{pmatrix}
& \mu_4 = \begin{pmatrix}
0 & 1 &0 \\
-1 & 0 &0 \\
0 & 0 &0
\end{pmatrix}.
\end{array}
\end{equation}
The general configuration of ${\bf d}_A$ and ${\bf d}_B$ can be obtained by operating the general rotation on the chosen ground state configurations ${\bf d}^{gs}_A$ and ${\bf d}^{gs}_B$ described above,
\begin{eqnarray}
{\bf D}(\phi, \ell, v) = e^{i \sum_{p=1}^4 \lambda_p \phi_p + i \mu_1 \ell^z_1 + i \mu_2 v_A + i \mu_3 v_B + i \mu_4 \ell_2^z},
\end{eqnarray}
where
\begin{eqnarray}
 {\bf d}_A = {\bf D}(\phi, \ell, v) {\bf d}^{gs}_A ,~~{\bf d}_B = {\bf D}(\phi, \ell, v) {\bf d}^{gs}_B.
\end{eqnarray}

Expanding the canting terms up to quadratic order, we can approximate
\begin{eqnarray}
{\bf d}_A \simeq {\bf U}(\phi) \begin{pmatrix}
1 - \frac{1}{2}\left[ (\ell^z_1)^2 + (\ell^z_2)^2 + v_A^2\right] \\
\ell^z_1 - i \ell^z_2 \\
i v_A
\end{pmatrix} \label{Eq:d_A}\\
{\bf d}_B \simeq {\bf U}(\phi) \begin{pmatrix}
\ell^z_1 + i \ell^z_2 \\
1 - \frac{1}{2}\left[ (\ell^z_1)^2 + (\ell^z_2)^2 + v_B^2\right] \\
-i v_B 
\end{pmatrix}\label{Eq:d_B}
\end{eqnarray}
Introducing $\ell^z \equiv \ell_1^z + i \ell^z_2$ and reparametrizing
\begin{eqnarray}
{\bf U} = \begin{pmatrix}
n^x_A & n^x_B & n^x_C \\
n^y_A & n^y_B & n^y_C \\
n^z_A & n^z_B & n^z_C 
\end{pmatrix}
\end{eqnarray}
where the vectors ${\bf n}_a$ inherit the constraints of ${\bf d}$ with
\begin{eqnarray}
&& {\bf n}_a \cdot \bar{\bf n}_{b\not=a} = 0,~~{\bf n}_a^2 = {\bf n}_a^2 \\
&& {\bf n}_a \cdot \bar{\bf n}_a = 1,
\end{eqnarray}
where the vector ${\bf n}_c \equiv \bar{\bf n}_A \times \bar{\bf n}_B$ is introduced as a convenient piece of book-keeping in the present $(\pi, \pi)$-AFQ (two sub lattice order), and is not an independent degrees of freedom.

Going to continuum limit involves the assumption that physically interesting variation takes place on a length scale much larger than the lattice constant $a \equiv 1$ and so gradients within the placate are small. In addition, the continuum field theory should describe the dynamics of both the broken symmetry state and the nearby paramagnetic region, in which the order parameter is assumed to be locally robust but slowly varying over macroscopic length sales. It is therefore necessary to allow the fields to fluctuate in space and time. From now on , all the parameters are allowed to fluctuate in space-time domain, $({\bf r}, \tau)$.

The partition function is
\begin{eqnarray}
\mathcal{Z}^{SU(3)}_\square = \int D[{\bf d}] e^{- S^{SU(3)}_\square [{\bf d}]},
\end{eqnarray}
where the action includes $S^{SU(3)}_\square = S_{kinetic} + S_{Hs}$. Let's focus on the Hamiltonian term first.

\subsection{Continuum theory for Hamiltonian terms }
The action for the Hamiltonian term is
\begin{eqnarray}
\nonumber S_{Hs} &=& \int_0^\beta d\tau H^{SU(3)}_{S} \\
& =& \frac{1}{2}\int_0^\beta d\tau \int d^2{\bf r} H^{SU(3)}_{S, clus},
\end{eqnarray}
where we focus on one cluster instead of a single site and the $2$ in the denominator is the effective cluster area. Within the cluster picture, we can write down the Hamiltonian term in terms of ${\bf d}_j$. Then we can perform a gradient expansion,
\begin{eqnarray}
\nonumber {\bf d}_j({\bf r} + {\bf \epsilon}_i, \tau) \simeq {\bf d}_j ({\bf r}, \tau) + ({\bf \epsilon}_i \cdot {\bf \nabla}){\bf d}_j({\bf r}, \tau) + \frac{1}{2} ( {\bf \epsilon}_i \cdot {\bf \nabla})^2{\bf d}_j({\bf r}, \tau),\\
\end{eqnarray} 
to give a continuum theory of the Hamiltonian term. For a square cluster with 4 sites, let's explicitly write down the Hamiltonian terms below.
\begin{eqnarray}
\nonumber H^{SU(3)}_{S,clus} = H^{SU(3)}_{J_1,clus} + H^{SU(3)}_{J_2, clus} , 
\end{eqnarray}
where
\begin{eqnarray}
\nonumber H^{SU(3)}_{J_1,clus} = && -J_1 \bigg{[} \left|{\bf d}_A ({\bf r} + (-\frac{1}{2}, -\frac{1}{2}), \tau) \cdot \bar{\bf d}_B ({\bf r} + (\frac{1}{2}, -\frac{1}{2}), \tau)\right|^2 + \left|{\bf d}_A ({\bf r} + (-\frac{1}{2}, -\frac{1}{2}), \tau)\cdot \bar{\bf d}_A({\bf r} + (-\frac{1}{2}, \frac{1}{2}), \tau)\right|^2 \\
&& + \left| {\bf d}_B({\bf r} + (\frac{1}{2},-\frac{1}{2}),\tau) \cdot \bar{\bf d}_A({\bf r} + (\frac{1}{2},\frac{1}{2}),\tau)\right|^2 + \left| {\bf d}_B({\bf r} + (-\frac{1}{2},\frac{1}{2}),\tau)\cdot \bar{\bf d}_A({\bf r}+(\frac{1}{2},\frac{2}{2}), \tau)\right|^2 \bigg{]}, \\
\nonumber H^{SU(3)}_{J_2, clus} = && J_2 \bigg{[}\left|{\bf d}_A({\bf r} + (-\frac{1}{2},-\frac{1}{2}),\tau)\cdot \bar{\bf d}_A({\bf r}+(\frac{1}{2},\frac{1}{2}),\tau)\right|^2 + \left| {\bf d}_B({\bf r} + (\frac{1}{2},-\frac{1}{2}),\tau)\cdot \bar{\bf d}_A({\bf r} + (-\frac{1}{2},\frac{1}{2}),\tau)\right|^2 \\
&& + \left| {\bf d}_A({\bf r} + (\frac{1}{2},\frac{1}{2}),\tau)\cdot \bar{\bf d}_A({\bf r} + (-\frac{1}{2},\frac{3}{2}),\tau)\right|^2 + \left| {\bf d}_B({\bf r}+(-\frac{1}{2},\frac{1}{2}),\tau) \cdot \bar{\bf d}_A({\bf r} + (\frac{1}{2}, \frac{3}{2}),\tau)\right|^2  \bigg{]}.
\end{eqnarray}

We now perform gradient expansion for each term
\begin{enumerate}
\item[(1)] $J_1$ terms: 
\begin{eqnarray}
 4 \left| {\bf d}_A \cdot \bar{\bf d}_B \right|^2 + 2\sum_{\lambda=x,y} \left| \bm{d}_A \cdot \partial_\lambda \bar{\bm{d}}_B\right|^2.
\end{eqnarray}

\item[(2)] $J_2$ terms: 
\begin{eqnarray}
-2 \sum_{\lambda = x, y} \left[ \left| \partial_\lambda \bm{d}_A \right|^2 + \left| \partial_\lambda \bm{d}_B \right|^2 - 2 \left| \bm{d}_A \partial_\lambda \bar{\bm{d}}_A \right|^2 - \left| \bm{d}_B \partial_\lambda \bar{\bm{d}}_B \right|^2 \right].
\end{eqnarray}
\end{enumerate}

Combining all terms leads and using Eqs.~(\ref{Eq:d_A})-(\ref{Eq:d_B}), 
\begin{eqnarray}
&& \left| {\bf d}_A \cdot\bar{\bf d}_B \right|^2 \simeq 4 \left[ \left( \ell^z_1\right)^2 + \left( \ell^z_2\right)^2 \right],\\
&& \left| {\bf d}_A \cdot \partial_{\lambda=x, y} \bar{\bf d}_B \right|^2 \simeq \left| {\bf n}_A \cdot \partial_{\lambda} \bar{\bf n}_B \right|^2,\\
&& {\bf d}_A \cdot \bar{\bf d}_A \simeq {\bf d}_B \cdot \bar{\bf d}_B \simeq 1,
\end{eqnarray}
we obtain that
\begin{eqnarray}
\nonumber H^{SU(3)}_{S, clus}\simeq &&16 J_1 \left[ \left( \ell^z_1 \right)^2 + \left( \ell^z_2 \right)^2 \right] + 2 J_1 \sum_{\lambda=x, y} \left| \bm{n}_A \partial_x \bar{\bm{n}}_B \right|^2 - 2 J_2 \sum_{\lambda= x,y} \bigg{[} \left| \partial_\lambda \bm{n}_A \right|^2 + \left| \partial_\lambda \bm{n}_B \right|^2 - \left| \bm{n}_A \partial_\lambda \bm{n}_A \right|^2 - \left| \bm{n}_B \partial_\lambda \bar{\bm{n}}_B\right|^2\bigg{]},\\ \label{Eq:SU(3)Hamil}
\end{eqnarray}
where we explicitly ignore constant terms.\\

\subsection{Continuum theory for the kinetic terms}
The action for the kinetic terms is quantum-mechanical in origin and is
\begin{eqnarray}
S_{kin} \simeq \int_0^\beta \frac{1}{2} \int d^2{\bf r} \sum_{a = A,B} \bar{\bf d}_a \partial_\tau {\bf d}_a = \frac{1}{2}\int_0^\beta \int d^2{\bf r} \left[ \bar{\bf d}_A \partial_\tau {\bf d}_A + \bar{\bf d}_B \partial_\tau {\bf d}_B\right].
\end{eqnarray}
Using again Eqs.~(\ref{Eq:d_A})-(\ref{Eq:d_B}) leads to the Lagrangian for the kinetic term as
\begin{widetext}
\begin{eqnarray}
\nonumber \mathcal{L}_{kin,clus} \simeq && \bar{\bf n}_A \partial_\tau {\bf n}_A + \bar{\bf n}_B \partial_\tau {\bf n}_B + 2 \ell^z_1 \left( \bar{\bf n}_A \partial_\tau {\bf n}_B - {\bf n}_A\partial_\tau \bar{\bf n}_B \right)  - 2 i \ell^z_2\left( \bar{\bf n}_A \partial_\tau {\bf n}_B + {\bf n}_A \partial_\tau \bar{\bf n}_B \right)\\
&& -i v_A \left( \bar{\bf n}_c \partial_\tau {\bf n}_A + {\bf n}_c \partial_\tau \bar{\bf n}_A \right) - i v_B \left( \bar{\bf n}_B \partial_\tau {\bf n}_c + {\bf n}_B \partial_\tau \bar{\bf n}_c \right).\label{Eq:SU(3)kinetic}
\end{eqnarray}
\end{widetext}
Therefore, the Lagrangian for continuum theory is $\mathcal{L}^{NLsM}_S = \mathcal{L}_{kin,clus} + \mathcal{L}^{SU(3)}_{S, clus}$, Eq.~\eqref{Eq:SU(3)Hamil} + Eq.~\eqref{Eq:SU(3)kinetic}. 

\subsection{Away from SU(3) point}
We now introduce SU(3) breaking terms to break the SU(3) down to SU(2). In principle, if we focus on the AFQ phase, the SU(2) symmetry will guarantee there are 3 gapless Goldstone modes associated with the quadrupole wave fluctuations. The SU(3)-breaking terms added are
\begin{eqnarray}
H' =  J'_1\sum_{\la i j \ra} \left( \vec{S}_i \cdot \vec{S}_j \right)^2 + J'_2\sum_{\la \la i j \ra \ra} \left( \vec{S}_i \cdot \vec{S}_j \right)^2.
\end{eqnarray}
In terms of ${\bf d}$ framework, 
\begin{eqnarray}
\nonumber H' = && J'_1 \sum_{\la i j\ra} \left| {\bf d}_i \cdot {\bf d}_j \right|^2 + J'_2 \sum_{\la\la i j \ra\ra}\left| {\bf d}_i \cdot {\bf d}_j \right|^2,
\end{eqnarray}
where we ignore the constant terms. We again focus on a cluster and perform gradient expansion. Let's write down each contributions separately.
\begin{enumerate}
\item[(1)] $J'_1$ terms:\\
\begin{eqnarray}
4\left| {\bf d}_A \cdot {\bf d}_B \right|^2 + 2 \sum_{\lambda= x,y} \left|{\bf d}_A \cdot\partial_\lambda {\bf d}_B\right|^2.
\end{eqnarray}

\item[(2)] $J'_2$ terms: \\
\begin{eqnarray}
 \bm{d}^2_A\bar{\bm{d}}_A^2 + \bm{d}^2_B \bar{\bm{d}}_B^2 - \sum_{\lambda = x, y} \left[ \left( \partial_\lambda \bm{d}_A\right)^2 + \left( \partial_\lambda \bar{\bm{d}}_A\right)^2 + \left( \partial_\lambda \bm{d}_B\right)^2 + \left( \partial_\lambda \bar{\bm{d}}_B \right)^2 \right].
\end{eqnarray}
\end{enumerate}

We then use Eqs.~(\ref{Eq:d_A})-(\ref{Eq:d_B}) simplify the results above in terms of the parametrization fields ${\bf n}_a$ and the canting fields $\ell^z_1, \ell^z_2, v_A, v_B$. The only new assumption we need to use is that in the AFQ phase the parametrization fields ${\bf n}_a$ inherit the constraints of the original vector ${\bf d}$. In AFQ, ${\bf d}$ is purely real or purely imaginary. Without loss of generality, we assume it is purely real. The additional constraints are 
\begin{eqnarray}
&& {\bf n}_a \cdot \bar{\bf n}_a = 1 \rightarrow {\bf n}_a^2 \simeq 1,\\
&& {\bf n}_a \cdot \bar{\bf n}_{b\not= a} = 0 \rightarrow {\bf n}_a \cdot {\bf n}_{b\not=a} \ll 1.
\end{eqnarray}
We can then obtain approximately,
\begin{eqnarray}
&& \bm{d}^2_A \simeq  1 - 2 \left( \ell^z_2 \right)^2 - 2 v_A^2,\\
&& \left| {\bf d}_A \cdot {\bf d}_A\right|^2 =\bm{d}^2_A \bar{\bm{d}}_A^2 \simeq 1 - 4 \left(\ell^z_2\right)^2 - 4 v_A^2,\\
&& \left| {\bf d}_B \cdot {\bf d}_B\right|^2 =\bm{d}_B^2 \bar{\bm{d}}_B^2 \simeq 1 - 4 \left(\ell^z_2\right)^2 - 4 v_B^2,
\end{eqnarray}
where we used the approximation $\bm{n}_A^2 \bar{\bm{n}}_A^2 \simeq 1 \simeq \bm{n}_B^2 \bar{\bm{n}}_B^2$. In the end, we can conclude that SU(3) breaking Hamiltonian i (ignoring constants)
\begin{widetext}
\begin{eqnarray}
\nonumber  H'_{clus} \simeq && J_1'\left[ 4 \left| \bm{n}_A \cdot \bm{n}_B \right|^2 + 16 \left( \ell^z_1\right)^2 + 2 \sum_\lambda \left| \bm{n}_A \partial_\lambda \bm{n}_B \right|^2\right] \\
&& - J_2' \left[ 16 \left( \ell^z_2 \right)^2 + 8 \left( v_A^2 + v_B^2\right) + \sum_\lambda \left( \left( \partial_\lambda \bm{n}_A \right)^2 + \left( \partial_\lambda \bar{\bm{n}}_A\right)^2 + \left( \partial_\lambda \bm{n}_B\right)^2 + \left( \partial_\lambda \bar{\bm{n}}_B\right)^2 \right) \right].\label{Eq:H'}
\end{eqnarray}
\end{widetext}
For obtaining the NLsM for the $(\pi,\pi)$-AFQ, we need to combine Eq.~(\ref{Eq:SU(3)Hamil}), Eq.~(\ref{Eq:SU(3)kinetic}), and Eq.~(\ref{Eq:H'}) and integrate out the canting fields within the steepest-decent approximation. Solving the differential equations,
\begin{equation}
\begin{array}{lr}
\frac{\delta \mathcal{L}}{\delta \ell^z} = 0; &~~~\frac{\delta \mathcal{L}}{\delta \ell^z} =0, \\
\frac{\delta \mathcal{L}}{\delta v_A} = 0; &~~~\frac{\delta \mathcal{L}}{\delta v_B} =0.
\end{array}
\end{equation}
We obtain 
\begin{eqnarray}
&& \ell^z_1 = - \frac{\bar{n}_A \partial_\tau n_B - n_A \partial_\tau \bar{n}_B}{16(J_1 + J_1')},\\
&& \ell^z_2 = \frac{i ( \bar{n}_A \partial_\tau n_B + n_A \partial_\tau \bar{n}_B)}{16(J_1 - J_2')},\\
&& v_A = - \frac{i ( \bar{n}_C \partial_\tau n_A + n_C \partial_\tau \bar{n}_A )}{16 J_2'},\\
&& v_B = - \frac{ i (\bar{n}_B \partial_\tau n_C + n_B \partial_\tau \bar{n}_C)}{16 J_2'}.
\end{eqnarray}
Plugging the solutions back to the Lagrangian, we obtain the $\mathcal{L}_{NLsM,plus}$ 

\begin{eqnarray}
\nonumber \mathcal{L}_{NLsM,clus} = && \bar{n}_A \partial_\tau n_A + \bar{n}_B \partial_\tau n_B - \\
\nonumber && - \frac{\left[ \bar{n}_A \partial_\tau n_B - n_A \partial_\tau \bar{n}_B\right]^2}{16 (J_1 + J_1')} + \frac{\left[ \bar{n}_A \partial_\tau n_B + n_A \partial_\tau \bar{n}_B \right]^2}{16(J_1 - J'_2)} - \frac{\left[ \bar{n}_C \partial_\tau n_A + n_C \partial_\tau \bar{n}_A \right]}{32 J_2'} - \frac{\left[ \bar{n}_B \partial_\tau n_C + n_B \partial_\tau \bar{n}_C\right]^2}{32 J_2'}+ \\
\nonumber && + 2 J_1 \sum_{\lambda = x, y} \left| n_A \partial_\lambda \bar{n}_B \right|^2 + 2 J_1'\sum_\lambda \left| n_A \partial_\lambda n_B \right|^2 - 2 J_2 \sum_\lambda \left[ \left| \partial_\lambda n_A \right|^2 + \left| \partial_\lambda n_B \right|^2 - \left| \bar{n}_A \partial_\lambda n_A\right|^2 - \left| \bar{n}_B \partial_\lambda n_B \right|^2 \right] + \\
 && - J_2' \sum_\lambda \left[ \left( \partial_\lambda n_A \right)^2 + \left( \partial_\lambda \bar{n}_A \right)^2 + \left( \partial_\lambda n_B \right)^2 + \left( \partial_\lambda \bar{n}_B\right)^2 \right] + 4 J_1' \left| n_A \cdot n_B \right|^2. 
\end{eqnarray}

We now derive the linearized action by re-expressing $\bm{n}_a$ in terms of bosonic fields $\phi_{1,2,3,4}$, with
\begin{eqnarray}\label{Eq:supp_n_to_phi}
\nonumber n_A \simeq \begin{pmatrix}
1 \\
-\phi_1 + i \phi_4 \\
\phi_2
\end{pmatrix};&~~
n_B \simeq \begin{pmatrix}
\phi_1 + i \phi_4 \\
1 \\
-\phi_3
\end{pmatrix};&~~
n_C \simeq \begin{pmatrix}
-\phi_2 \\
\phi_3 \\
1
\end{pmatrix},\\
\end{eqnarray}
the linearized harmonic Lagrangian for $(\pi,\pi)$-AFQ is
\begin{eqnarray}
\nonumber \mathcal{L} \simeq && \frac{ ( \partial_\tau \phi_1 )^2}{4(J_1 + J_2 - K_2)} + 2 (K_1 - 2 K_2) \sum_\lambda (\partial_\lambda \phi_1)^2  + \sum_{a=2,3} \left[ \frac{(\partial_\tau \phi_a)^2}{8( J_2 - K_2)} - 2K_2 \sum_\lambda (\partial_\lambda \phi_a)^2\right] + \\
\nonumber && + \frac{(\partial_\tau \phi_4)^2}{4 K_1} + 2 (K_1 +2 K_2 - 4J_2)\sum_\lambda (\partial_\lambda \phi_4)^2 + 16( K_1 - J_1)\phi_4^2 - \\
&& \equiv \mathcal{L}^{NLsM}_S(\phi_1, \phi_2, \phi_3, \phi_4).
\end{eqnarray}

Ignoring the terms thtat couple to the conduction electrons' degrees of freedom, we can see the stability for $(\pi,\pi)$-AFQ requires positive stiffness leading to the conditions
\begin{eqnarray}
&& K_1 >0,~~~~K_2 <0,\\
&& K_1 - J_1 \geq 0,~~~~ J_2-K_2 >0,\\
&& J_1 + J_2 - K_2 > 0,~~~~K_1 + 2K_2 - 4J_2 \geq 0.
\end{eqnarray}
If we focus on the regime where $K_1 > J_1$, we can see that $\phi_4$ is always gapped due to the finite mass term and can be ignored.  The effective low-energy description of the action, $S^{NLsM}_{(\pi,\pi)-AFQ}$, is
\begin{eqnarray}
\nonumber S^{NLsM}_{(\pi,\pi)-AFQ} &\simeq& \int_0^\beta \frac{d\tau}{2} \int d^2 r \Bigg{\{} \frac{ ( \partial_\tau \phi_1 )^2}{4(J_1 + J_2 - K_2)} + 2 (K_1 - 2 K_2) \sum_\lambda (\partial_\lambda \phi_1)^2 + \sum_{a=2,3} \left[ \frac{(\partial_\tau \phi_a)^2}{8( J_2 - K_2)} - 2K_2 \sum_\lambda (\partial_\lambda \phi_a)^2\right] \Bigg{\}}. \\
\end{eqnarray}

\section{Kondo effects in the $(\pi,\pi)$-AFQ}
We now consider the Kondo effects in the $(\pi,\pi)$-AFQ. Eqs.~\eqref{Eq:d_A}-\eqref{Eq:d_B} give the general parameterizations for the $\bm{d}$ vectors on the sublattices A and B as
\begin{eqnarray}
&& \bm{d}_A = \bm{n}_A \left[ 1 - \frac{1}{2} \left| \ell^z\right|^2 - \frac{1}{2} v_A^2 \right] + \bm{n}_B \bar{\ell}^z + \bm{n}_c \left( i v_A \right), \\
&& \bm{d}_B = \bm{n}_B \left[ 1 - \frac{1}{2} \left| \ell^z \right|^2 - \frac{1}{2} v_B^2 \right] + \bm{n}_A \ell^z + \bm{n}_c \left( - i v_B \right).
\end{eqnarray}
In order to extract the low-energy description of the Kondo coupling, we first need to extract the low-energy continuum theory descriptions of the 8 component operator that consists of 3-component $S^{x,y,z}$, and 5-component $Q^{x^2-y^2},~Q^{3z^2 - r^2},~Q^{xy},~Q^{yz},~Q^{zx}$. Since we have the expressions of the fluctuating $\bm{d}$ vectors, we can write down the fluctuating $8$-component dipole-quadrupole moment,
\begin{eqnarray}
\mathcal{Q}_A = \begin{pmatrix}
0\\
-2v_A +i(n^z_A - \bar{n}_A^z)\\
i \left(\ell^z - \bar{\ell}^z\right) - i (n^y_A - \bar{n}_A^y) \\
-1 \\
\frac{1}{\sqrt{3}}\\
-\left( \ell^z + \bar{\ell}^z \right)-(n^y_A + \bar{n}_A^y)\\
0\\
-(n^z_A + \bar{n}^z_A)\\
\end{pmatrix},&~~
\mathcal{Q}_B = \begin{pmatrix}
-2v_B - (n^z_B - \bar{n}^z_B)\\
0\\
i\left( \ell^z - \bar{\ell}^z\right) + i(n^x_B - \bar{n}^x_B)\\
1 \\
\frac{1}{\sqrt{3}}\\
-\left( \ell^z + \bar{\ell}^z \right)-(n^x_B + \bar{n}^x_B)\\
-(n^z_B + \bar{n}^z_B)\\
0
\end{pmatrix}.
\end{eqnarray}
We can then write down the low-energy descriptions of the 8-component moment as
\begin{eqnarray}
\mathcal{Q}(\bm{r},t) & \simeq & \begin{pmatrix}
-\left[v_B + \frac{i}{2}\left( n^z_B - \bar{n}^z_B\right) \right] \\
-\left[ v_A -\frac{i}{2} \left( n^z_A - \bar{n}^z_A \right)\right] \\
-2 \ell^z_2-\frac{i}{2} \left( n^y_A - \bar{n}^y_A - n^x_B + \bar{n}^x_B \right) \\
0\\
\frac{1}{\sqrt{3}}\\
-2 \ell^z_1 -\frac{1}{2} \left( n^y_A + \bar{n}^y_A + n^x_B + \bar{n}^x_B \right)\\
-\frac{1}{2}\left( n^z_B + \bar{n}_B^z\right)\\
-\frac{1}{2} \left( n^z_A + \bar{n}^z_A \right)
\end{pmatrix}
+ \begin{pmatrix}
-\left[ v_B +\frac{i}{2}\left( n^z_B - \bar{n}^z_B \right)\right] \\
\left[ v_A - \frac{i}{2}\left( n^z_A - \bar{n}^z_A\right)\right]  \\
\frac{i}{2}\left( n^y_A - \bar{n}^y_A + n^x_B - \bar{n}^x_B \right)\\
1\\
0\\
\frac{1}{2}\left[ n^y_A + \bar{n}^y_A - \left( n^x_B + \bar{n}^x_B\right)\right]\\
-\frac{1}{2}\left( n^z_B + \bar{n}^z_B\right)\\
\frac{1}{2}\left( n^z_A + \bar{n}^z_A \right) 
\end{pmatrix}\cos(\bm{K}\cdot \bm{r}) \\
& \equiv & \bm{Q}_{\bm{0}} + \bm{Q}_{\bm{M}} \cos\left( \bm{M}\cdot \bm{r}\right),
\end{eqnarray}
where we explicitly separate out the uniform part and the oscillatory part. First we note that the uniform part contains a ``static" background of $Q^{3z^2 -r^2}$ that directly couple to the $q_c^{3z^2 -r^2}$ of the 3-flavor conduction electrons. This static background $Q^{3z^2 - r^2}$ field is only invariant under rotation between $x$-$y$ plane, which, therefore, should break the SU(3) symmetry of the conduction elections dow to SU(2)$\times$U(1), where SU(2) is spanned by the $c_x$ and $c_y$ conduction elections and U(1) is spanned by $c_z$. We then expect that the low-energy descriptions of conduction electrons $c_x$ and $c_y$ should be different from that of $c_z$, \textit{i.e.}, the Fermi velocities of $c_x$ and $c_y$ are the same ($v_x = v_y$) but is different from that of $c_z~(v_z)$. 

Explicitly, the low-energy theory of the conduction electrons in the presence of the static quadrupolar background is
\begin{eqnarray}
S_c \simeq && \sum_{\alpha = x,y} \int d^d \bm{K} d\epsilon \psi^\dagger_\alpha (\bm{K},i\epsilon) (i\epsilon - \xi_k)\psi_\alpha(\bm{K},i\epsilon) + \int d^d \bm{K}' d\epsilon' \psi^\dagger_z (\bm{K}', i \epsilon')(i\epsilon' - \tilde{\xi}_{K'}) \psi_z (\bm{K}', i \epsilon'),
\end{eqnarray}
where $\xi_K = v_F(K - K_F)$, and $\tilde{\xi}_{k'} = \tilde{v}_{F}(K' - K'_F)$, where $k_F$ and $k'_F$ are generically different. We want to remark that due to the band splitting between the $x,y$- and $z$- fermions, the spin dipolar and quadrupolar degrees of freedom consisting of $\psi_{x/y}$ and $\psi_{z}$ fermions, $i.~e.$, $s^x_c \sim \psi^\dagger_y \psi_z - \psi^\dagger_z \psi_y$, can be ignored due to the finite energy gap.

Since we have the NLsM description of all the fields, we can also re-express the Kondo couplings within the NLsM construction. Ignoring the gapped fields due to the static $\la Q^{3z^2- r^2}\ra$ background in the $(\pi,\pi)$-AFQ, we can write down the Kondo coupling Lagrangian as
\begin{eqnarray}
 \mathcal{L}_K = && -J_K^I   \left( 2\ell^z_2 + \frac{i}{2} \left(n^y_A - \bar{n}^y_A - n^x_B  + \bar{n}^x_B\right)\right)s_c^z  - J_K^{II}  \left( 2 \ell^z_1 + \frac{1}{2}\left( n^y_A + \bar{n}^y_A + n^x_B + \bar{n}^x_B \right)\right) q_c^{xy} .~~~
\end{eqnarray}
We again integrate out the canting fields by utilizing the steepest descent approximation with
\begin{equation}
\begin{array}{lr}
\frac{\delta \mathcal{L}}{\delta \ell^z} = 0; &~~~\frac{\delta \mathcal{L}}{\delta \ell^z} =0, \\
\frac{\delta \mathcal{L}}{\delta v_A} = 0; &~~~\frac{\delta \mathcal{L}}{\delta v_B} =0.
\end{array}
\end{equation}
We then get the results
\begin{eqnarray}
&& \ell^z_1 = - \frac{\bar{n}_A \partial_\tau n_B - n_A \partial_\tau \bar{n}_B}{16(J_1 + J_1')} + \frac{J_K^{II}}{16(J_1 + J_1')} q_c^{xy},\\
&& \ell^z_2 = \frac{i ( \bar{n}_A \partial_\tau n_B + n_A \partial_\tau \bar{n}_B)}{16(J_1 - J_2')} + \frac{J^I_K}{16(J_1 - J'_2)} s^z_c,\\
&& v_A = - \frac{i ( \bar{n}_C \partial_\tau n_A + n_C \partial_\tau \bar{n}_A )}{16 J_2'},\\
&& v_B = - \frac{ i (\bar{n}_B \partial_\tau n_C + n_B \partial_\tau \bar{n}_C)}{16 J_2'}.
\end{eqnarray}
We can plug the solutions back to the Lagrangian to obtain
\begin{eqnarray}
\nonumber \mathcal{L} = && \bar{n}_A \partial_\tau n_A + \bar{n}_B \partial_\tau n_B - \\
\nonumber && - \frac{\left[ \bar{n}_A \partial_\tau n_B - n_A \partial_\tau \bar{n}_B\right]^2}{16 (J_1 + J_1')} + \frac{\left[ \bar{n}_A \partial_\tau n_B + n_A \partial_\tau \bar{n}_B \right]^2}{16(J_1 - J'_2)} - \frac{\left[ \bar{n}_C \partial_\tau n_A + n_C \partial_\tau \bar{n}_A \right]}{32 J_2'} - \frac{\left[ \bar{n}_B \partial_\tau n_C + n_B \partial_\tau \bar{n}_C\right]^2}{32 J_2'}+ \\
\nonumber && + 2 J_1 \sum_{\lambda = x, y} \left| n_A \partial_\lambda \bar{n}_B \right|^2 + 2 J_1'\sum_\lambda \left| n_A \partial_\lambda n_B \right|^2 - 2 J_2 \sum_\lambda \left[ \left| \partial_\lambda n_A \right|^2 + \left| \partial_\lambda n_B \right|^2 - \left| \bar{n}_A \partial_\lambda n_A\right|^2 - \left| \bar{n}_B \partial_\lambda n_B \right|^2 \right] + \\
\nonumber && - J_2' \sum_\lambda \left[ \left( \partial_\lambda n_A \right)^2 + \left( \partial_\lambda \bar{n}_A \right)^2 + \left( \partial_\lambda n_B \right)^2 + \left( \partial_\lambda \bar{n}_B\right)^2 \right] + 4 J_1' \left| n_A \cdot n_B \right|^2 -\\
\nonumber && - J_K^I \left[ \frac{i}{2}\left( n^y_A - \bar{n}^y_A - n^x_B + \bar{n}^x_B\right) + \frac{ i \left( \bar{n}_A \partial_\tau n_B + n_A \partial_\tau n_B \right)}{8(J_1 - J'_2)}\right] s_c^z  \\
\nonumber && - J_K^{II} \left[ \frac{1}{2}\left( n^y_A + \bar{n}^y_A + n_B^x + \bar{n}_B^x\right) - \frac{\bar{n}_A \partial_\tau n_B - n_A \partial_\tau \bar{n}_B}{8(J_1 + J'_1)}\right] q_c^{xy} - \\
 && - \frac{ (J_K^I )^2}{16(J_1 - J_2')} (s_c^z)^2 - \frac{(J_K^{II})^2}{16( J_1 + J_1')}(q_c^{xy})^2,
\end{eqnarray}
where we explicitly suppress the constant static backgroun $Q^{3z^2 - r^2}$ field which breaks SU(3) symmetry between three flavored conduction electrons. Focusing on the quadrupolar order, we again use the identity $\bar{n}_a \partial_\tau n_a = 1/2 \partial_\tau (n_a)^2 \simeq 0$, and can be ignored.

Re-express the fields in the bosonic fields using Eq.~\ref{Eq:supp_n_to_phi}, the Lagrangian becomes
\begin{eqnarray}
\nonumber \mathcal{L} \simeq && \frac{ ( \partial_\tau \phi_1 )^2}{4(J_1 + J_2 - K_2)} + 2 (K_1 - 2 K_2) \sum_\lambda (\partial_\lambda \phi_1)^2  + \sum_{a=2,3} \left[ \frac{(\partial_\tau \phi_a)^2}{8( J_2 - K_2)} - 2K_2 \sum_\lambda (\partial_\lambda \phi_a)^2\right] + \\
\nonumber && + \frac{(\partial_\tau \phi_4)^2}{4 K_1} + 2 (K_1 +2 K_2 - 4J_2)\sum_\lambda (\partial_\lambda \phi_4)^2 + 16( K_1 - J_1)\phi_4^2 - \\
\nonumber && - \frac{ i J^I_K}{4(J_1 + J_2 - K_2)}s^z_c (\partial_\tau \phi_1) + \frac{i J_K^{II}}{4K_1} q^{xy}_c (\partial_\tau \phi_4) - \frac{ (J_K^{II})^2}{16 K_1} (q^{xy}_c)^2- \frac{ (J_K^{I})^2}{16(J_1 + J_2 - K_2 )}(s^z_c)^2\\
&& \equiv \mathcal{L}^{NLsM}_S(\phi_1, \phi_2, \phi_3, \phi_4) + \mathcal{L}_K + \mathcal{L}(\psi^4),
\end{eqnarray}
where the last two terms $\mathcal{L}(\psi^4)$ only renormalize the Fermi velocity and the quartic fermion terms and can be ignored. Focusing on the point away from the SU(3) point, $K_1 >J_1$, we can also ignore $\phi_4$ due to the finite mass term. In the end we conclude the NLsM for the $(\pi,\pi)$-AFQ in the presence of Kondo couplings to the 3-flavor conduction electrons is 
\begin{eqnarray}
\nonumber S^{NLsM}_{(\pi,\pi)-AFQ} &\simeq& \int_0^\beta \frac{d\tau}{2} \int d^2 r \Bigg{\{} \frac{ ( \partial_\tau \phi_1 )^2}{4(J_1 + J_2 - K_2)} + 2 (K_1 - 2 K_2) \sum_\lambda (\partial_\lambda \phi_1)^2  + \sum_{a=2,3} \left[ \frac{(\partial_\tau \phi_a)^2}{8( J_2 - K_2)} - 2K_2 \sum_\lambda (\partial_\lambda \phi_a)^2\right] \Bigg{\}},\\ \label{Eq:supp_Ls}\\
S_K & \simeq & \int_0^\beta \frac{d\tau}{2} \int d^2 r \left( \lambda_z s^z_c \partial_\tau \phi_1 \right),\\
S_c & \simeq & \int d^d\bm{K} d\epsilon \sum_{\alpha = x,y} \psi^\dagger_\alpha (\bm{K}, i \epsilon) ( i \epsilon - \xi_K)\psi_\alpha(\bm{K},i\epsilon) + \int d^d\bm{K}' d\epsilon' \psi^\dagger_z(\bm{K}',i\epsilon')(i\epsilon' - \tilde{\xi}_{K'}) \psi_z(\bm{K}',i\epsilon'),
\end{eqnarray}
where we define  $\lambda_z = -i J_K^I/[4(J_1 + J_2 - K_2 )]$, $\xi_K = v_F(K-K_F)$, and $\tilde{\xi}_{K'} = \tilde{v}_F(K' - \tilde{K}'_F)$.

\section{Exact marginality of Kondo couplings}
In this work, we follow the scaling procedure illustrated in Ref.~\cite{Yamamoto2010} to conclude the exact marginality of the Kondo coupling $\lambda_z$ in the $(\pi,\pi)$-AFQ. For clarity in the scaling analysis, we first define $\Phi_a(\bm{r},\tau) \equiv \partial_\tau \phi_a(\bm{r},\tau)$. The scaling dimension of $\Phi_a(\bm{r},\tau)$ can be directly read out, $\Delta[\Phi_a(\bm{r},\tau)] = 1$, which leads to the scaling dimension of its Fourier partner as $\Delta[\Phi_a (\bm{q},\omega)] = -d$, where $d$ is the spatial dimension. For the conduction electron fields, the scaling dimension of a sermonic field is $\Delta[\psi_c(\bm{K}, \omega)] = -3/2$. We can see that at tree level, the scaling dimension of the Kondo coupling is \textit{marginal}, 
\begin{eqnarray}
\Delta[S_K] \bigg{|}_{tree-level} = \Delta[ dk d\epsilon d^dq d \omega \psi^{x \dagger}_c (\bm{k} + \bm{q}, \epsilon + \omega) \psi^y_c (\bm{k}, \omega) \Phi_1 (\bm{q},\omega)] = 1 + 1 + d + 1 + 2 (-3/2) -d = 0.
\end{eqnarray} 
Now we will show that the result at tree level is exact based on procedure in Ref.~\cite{Yamamoto2010}.

Considering a spherical Fermi surface of conduction electrons, we first re-express the momentum integral in the part of the action of the conduction electron as the momentum integral near the Fermi surface. Keeping the most relevant term, we obtain 
\begin{eqnarray}
\int d^d \bm{K} = \int^{K_F + \Lambda}_{K_F - \Lambda} K^{d-1} dK \int d^{d-1} \Omega_K \simeq  K_F^{d-1} \int_{-\Lambda}^\Lambda dk \int d^{d-1} \Omega_K
\end{eqnarray}
, where we introduce $ k = K_F - K$ and keep only the $K_F^{d-1}$ terms after Taylor expansion. Now the kinetic part of the fermions can be re-expressed as
\begin{eqnarray}
S_c \simeq K_F^{d-1} \int dk_\alpha d^{d-1} \Omega_K d\epsilon \psi^{\alpha \dagger}_c \left( i \epsilon_\alpha - v_F^\alpha k_\alpha \right) \psi^\alpha_c = \int d\bar{k}_\alpha d^{d-1}\bar{\Omega}_K d\bar{\epsilon} \bar{\psi}_c^{\alpha \dagger} \left ( \bar{\epsilon}_\alpha - v_F^\alpha \bar{k}_\alpha \right) \bar{\psi}_c^\alpha,
\end{eqnarray}
where we introduce the dimensionless couplings, $\epsilon = \Lambda \bar{\epsilon}$, $k = \Lambda \bar{k}$, $\Omega_K = \bar{\Omega}_K$, $K_F^{d-1} \Lambda^3 \psi^\dagger \psi = \bar{\psi}^\dagger \bar{\psi}$. For the action of the bosonic fields, Eq.~\eqref{Eq:supp_Ls}, we perform similar transformation, $\int d^2\bm{q} d\omega (\bm{q}^2 + \omega^2) \phi_1^2(\bm{q},\omega) = \int d^2\bar{\bm{q}} d \bar{\omega} ( \bar{\bm{q}}^2 + \bar{\omega}^2 ) \bar{\phi}_1^2 (\bar{\bm{q}},\bar{\omega})$, where we define $\Lambda^5 \phi_1^2 = \bar{\phi}_1^2$. Plugging the new definition into the Kondo action, we find that at $d=2$ it takes the form $\sim \int dk d\epsilon d^2q d\omega \psi_c^{x \dagger} (k+q, \epsilon +\omega)   \psi_c^y (k,\omega)  \omega  \phi_1 (q, \omega)$, which leads to
\begin{eqnarray}
\frac{S_K}{S_c} \propto  \Lambda^{1 + 1 + 2 + 1} \Lambda^{-3} K_F^{1-2} \Lambda^{1 - \frac{5}{2}} = \frac{\sqrt{\Lambda}}{K_F},
\end{eqnarray}
where we can see in the limit $\Lambda/K_F \rightarrow 0$, \textit{i.e.}, the Fermi momentum is much larger than the thin-shell momentum cut-off near the Fermi surface, the Kondo coupling is heavily suppressed.Therefore, the Kondo vertex is associated with positive powers of $\sqrt{\Lambda}/K_F$ which is vanishingly small. As the number of powers of Kondo couplings increases, so does the suppression factor, and, therefore, all higher-order terms are suppressed, which means that the scaling result at tree-level RG analysis is exact. The Kondo coupling is indeed exactly \textit{marginal}.

\bibliography{biblio4AFQ}